\documentclass[12pt]{emulateapj}

\usepackage[utf8]{inputenc}
\usepackage[usenames,dvipsnames]{color}
\usepackage{natbib}
\usepackage{graphicx}
\usepackage{comment}
\usepackage{amsbsy}
\usepackage{amsmath}
\usepackage{float}
\usepackage{amsmath}

\begin{document}

\title{Extragalactic Imprints in Galactic Dust Maps}

\shorttitle{Dust Map Tomography}
\shortauthors{Chiang \& M{\'e}nard}

\author{
Yi-Kuan Chiang\altaffilmark{1} and Brice M{\'e}nard\altaffilmark{1,2}
}
\altaffiltext{1}{Department of Physics \& Astronomy, Johns Hopkins University, 3400 N. Charles Street, Baltimore, MD 21218, USA}
\altaffiltext{2}{Kavli Institute for the Physics and Mathematics of the Universe, University of Tokyo, Kashiwa 277-8583, Japan}


\begin{abstract}
Extragalactic astronomy relies on the accurate estimation of source photometry corrected for Milky Way dust extinction. This has motivated the creation of a number of ``Galactic" dust maps. We investigate whether these maps are contaminated by extragalactic signals using the clustering-redshift technique, i.e., by measuring a set of angular cross-correlations with spectroscopic objects as a function of redshift. Our tomographic analysis reveals imprints of extragalactic large-scale structure patterns in nine out of 10 Galactic dust maps, including all infrared-based maps as well as ``stellar'' reddening maps. When such maps are used for extinction corrections, this extragalactic contamination introduces redshift- and scale-dependent biases in photometric estimates at the millimagnitude level. It can affect both object-based analyses, such as the estimation of the Hubble diagram with supernovae, as well as spatial statistics. The bias can be appreciable when measuring angular correlation functions with low amplitudes, such as lensing-induced correlations or angular correlations for sources distributed over a broad redshift range. As expected, we do not detect any extragalactic contamination for the dust map inferred from 21cm HI observations. Such a map provides an alternative to widely used infrared-based maps but relies on the assumption of a constant dust-to-gas ratio. We note that, using the WISE 12 micron map sensitive to polycyclic aromatic hydrocarbons (PAH), an indirect dust tracer, we detect the diffuse extragalactic PAH background up to $z\sim2$. Finally, we provide a procedure to minimize the level of biased magnitude corrections in maps with extragalactic imprints.
\end{abstract}


\section{Introduction}\label{sec:intro}

\begin{figure*}[t]
    \centering 
         \includegraphics[width=0.71\textwidth]{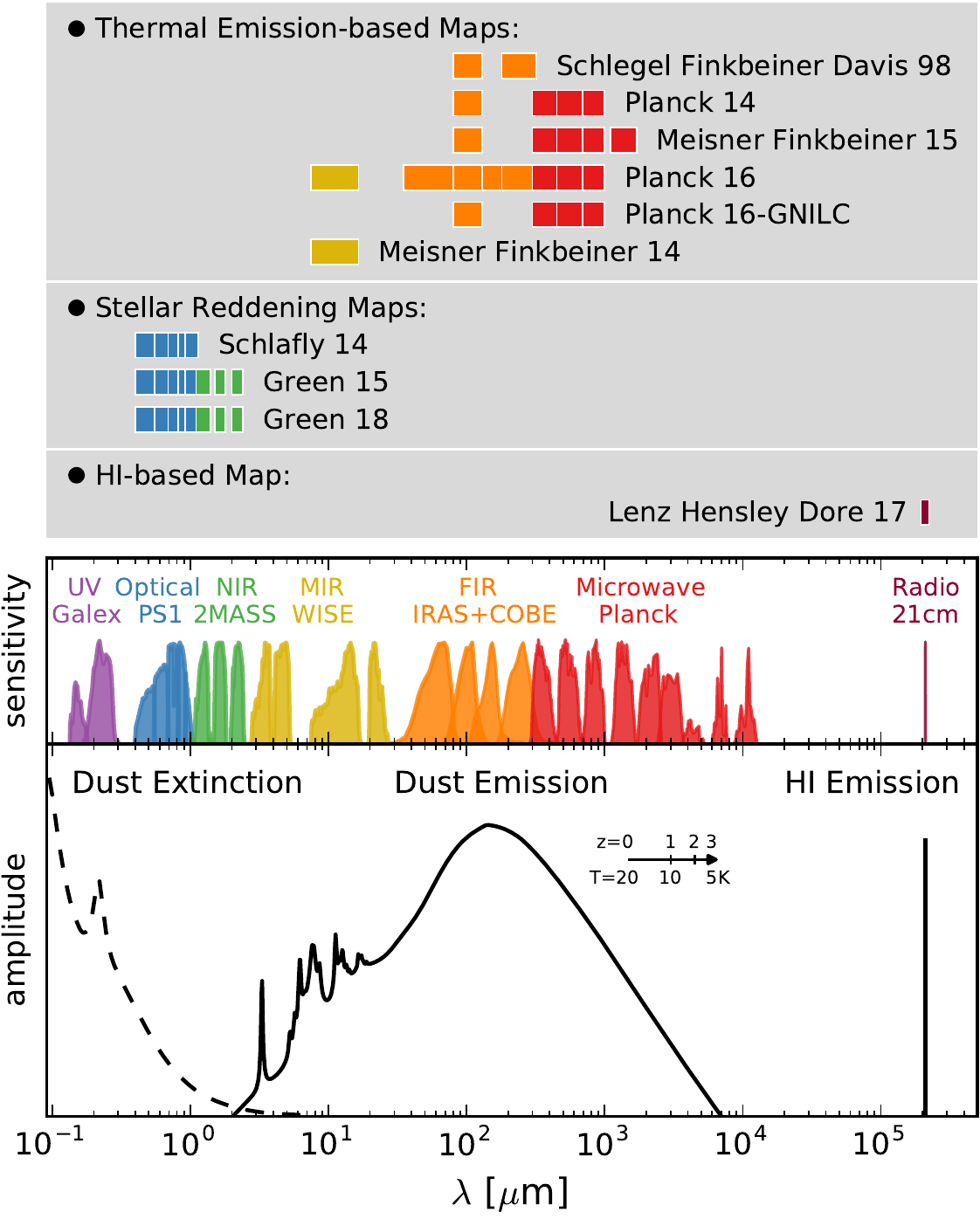}
         \caption{
         Bottom panel: spectral features of dust extinction \citep[dashed line; $A_{\lambda}$ from][]{2001ApJ...548..296W} and emission \citep[solid line; ${\rm log}\ j_{\nu}$ from][]{2007ApJ...657..810D}, together with a delta function for HI 21cm emission. The arrow shows the expected shift when emission temperature or redshift changes. A set of filter curves from several surveys is shown above with arbitrary normalization. Top panel: summary of spectral sampling for the 10 Galactic dust-reddening maps considered in this paper.
         }
\label{fig:dust_observables}
\end{figure*}

Our view of the extragalactic sky is altered by the presence of dust in the interstellar medium (ISM) of the Milky Way. Dust grains absorb and scatter incident photons at short wavelengths and emit radiation at longer wavelengths. Collectively, they produce a foreground screen that extincts and reddens the light of extragalactic objects. Correcting for such extinction effects is important for the accurate estimation of extragalactic source photometry, and the need for higher accuracy keeps increasing from the demanding requirements of precision cosmological experiments.

Dust can be traced over a wide range of wavelengths. This is illustrated in Figure~\ref{fig:dust_observables} where we show the expected extinction by Milky Way-type dust in the ultraviolet to near-infrared (IR) from \cite{2001ApJ...548..296W}, and the dust thermal emission from the model of \cite{2007ApJ...657..810D} with a broad peak in the far-IR and polycyclic aromatic hydrocarbon (PAH) features in the mid-IR. As dust is usually mixed with neutral gas, we also show the 21cm hydrogen emission line, which has been used as an indirect dust tracer. This latter technique was used to introduce the first large-scale dust map by
\cite{1978ApJ...225...40B,1982AJ.....87.1165B}. Later, the $IRAS$ and $COBE$ satellites launched in the 1980s opened up the atmospherically opaque window in the mid- to far-IR. \citet[][hereafter SFD]{1998ApJ...500..525S} used this data to create a dust-reddening map based on the thermal continuum emission of dust. Additional dust maps have been derived using IR data from the $Planck$ satellite,  PAH emission using the Wide-field Infrared Survey
Explorer ($WISE$), and optical reddening toward background stars \citep{2012ApJ...757..166B} or galaxies \citep{2010ApJ...719..415P}. The tight correlation between dust reddening and some of the diffuse interstellar bands has also been demonstrated and used to map Galactic dust in the optical \citep{2015MNRAS.452.3629L} and the near-IR \citep{2015ApJ...798...35Z}.

Due to the broad width of the blackbody spectrum, Galactic dust maps derived from IR emission measurements unavoidably include source contributions from extragalactic objects over a wide range of redshifts. Such maps therefore suffer from some level of extragalactic contamination. 

Such a signature has been reported by \cite{2007PASJ...59..205Y} who analyzed the number counts of low-redshift galaxies from the Sloan Digital Sky Survey (SDSS) as a function of the Galactic reddening measured in SFD. They showed that, toward low-reddening sight lines, the reddening value $E_{B-V}$ in SFD appears to be roughly proportional to the number density of galaxies, whereas it should not depend on such a quantity. Motivated by this result, \cite{2013PASJ...65...43K} showed that the mean reddening values derived by SFD at the locations of SDSS photometric galaxies and quasars are in excess with respect to their angular vicinity. This excess reddening can be largely accounted for by the contamination in far-IR emission from both the stacked galaxies and the galaxies clustered around them. \citet{2015MNRAS.446.2696S} applied the clustering-based redshift estimation technique to probe the extragalactic contribution of various intensity maps produced by the Planck Collaboration. These authors generalized the clustering-redshift technique (originally introduced for discrete sources by \citealt{2008ApJ...684...88N} and \citealt{2013arXiv1303.4722M}) to diffuse fields. This allowed them to reveal extragalactic contributions in the $Planck$ dust map.

In this paper, we follow a similar line of investigation. We analyze 10 Galactic dust maps available in the literature and perform a systematic clustering-redshift analysis for each of them in a uniform manner. We constrain the angular and redshift dependence of the extragalactic imprints in each dust map by measuring a set of angular cross-correlations between these maps with a spectroscopic reference sample of galaxies and quasars as a function of redshift. In Section~\ref{sec:formalism} we lay out the formalism of extinction correction, the impact of a biased reddening associated with extragalactic imprints on source number counts and clustering statistics, and how to measure it using cross-correlations. In Section~\ref{sec:data} we introduce the dust maps and cross-correlation reference sample. In Section~\ref{sec:exgalresults}, we present the results in measuring the amplitudes and redshift dependence of extragalactic imprints in each dust map. We discuss the implications for precision cosmology and summarize our results in Section~\ref{sec:discussion}, and \ref{sec:summary}, respectively. A flat universe with Planck cosmological parameters is assumed \citep{2014A&A...571A..16P}.

\section{Extinction correction and extragalactic contamination}\label{sec:formalism}

Here we introduce the formalism of extinction corrections and describe some of the biases that can be introduced in this process. Dust extinction toward a background object is commonly written as 
\begin{equation}
A_{\lambda} = m_{\rm obs}-m,
    \label{eq:total_ext}
\end{equation}
where $m_{\rm obs}$ and $m$ are the observed and intrinsic magnitudes evaluated at a given wavelength $\lambda$. 

Reddening is commonly measured using the optical $B$- and $V$-band color excess:
\begin{equation}
    E_{B-V} = A_B - A_V\,.
\end{equation}
The wavelength dependence of extinction can be parameterized using the total-to-selective extinction ratio
\begin{equation}
    R_{\lambda}\equiv A_{\lambda}/E_{B-V}\;.
    \label{eq:RV}
\end{equation}
The extinction curves $R_{\lambda}$ are generally nonlinear but can be approximately described by a family of curves using one parameter, often chosen to be $R_V$ \citep{1989ApJ...345..245C}. Precise characterizations of $R_{\lambda}$ in the UV require additional parameters. For diffuse ISM in the Milky Way, $R_V$ is about $3.1$ \citep{1975A&A....43..133S,2016ApJ...821...78S}.

To estimate the dereddened (or true) magnitude $m_{\rm dered}$ of an extragalactic object at an angular position $\phi$, one needs an estimate of the Galactic dust extinction,
\begin{eqnarray}
    \hat{m}_{\rm dered}(\phi) &=& m_{\rm obs}(\phi) - \hat A_{\lambda}(\phi)\, ,
    \label{eq:ext_correction_1}
\end{eqnarray}
where $\hat A_{\lambda}(\phi)$ is provided by a given dust-reddening map assuming an extinction curve. We can expect this quantity to have several components,
\begin{equation}
{\hat A}_{\lambda}(\phi) = A_{\lambda}^{\rm G}(\phi)+\delta A_{\lambda}^{\rm EG}(\phi) +\epsilon(\phi) 
    \label{eq:ext_components}
\end{equation}
where $A_{\lambda}^{\rm G}(\phi)$ and $\delta A_{\lambda}^{\rm EG}(\phi)$ are the true Galactic extinction and the bias due to any potential extragalactic imprints and $\epsilon(\phi)$ is the noise associated with the map estimate. Comparing the estimated and the true dereddened magnitude, we expect a magnitude shift $\delta m$, such that

\begin{eqnarray}
    \hat{m}_{\rm dered}\, &=&\,  m_{\rm dered} + \delta m;\\
    \delta m(\phi)\, &=&\, - \delta A_{\lambda}^{\rm EG}(\phi) - \epsilon(\phi)\, ,  
    \label{eq:ext_overcorrection}
\end{eqnarray}
where $\delta m$ is negative if the extinction is overestimated. This equation shows that the choice of a dust map should consider both its noise level (precision) and potential biases due to extragalactic sources (accuracy). When estimating the mean magnitude of an extragalactic source population, the noise term is expected to be statistically reduced, but the extragalactic imprints in the dust map behave as a systematic effect that introduces a correlated bias,
\begin{equation}
    \langle \delta m \rangle_{g} = 
    - \langle  \delta A_{\lambda}^{\rm EG} \rangle_{g}\;,
    \label{eq:pop_mag_shift}
\end{equation}
where the subscript $g$ refers to the positions of these extragalactic objects. As we will see, IR-based dust maps tend to overpredict Galactic dust extinction toward extragalactic objects. Extinction corrections using such maps thus tend to overestimate their brightness.

\subsection{Number Counts and Clustering Biases}\label{sec:discussionMagnification}
\label{sec:number_counts_and_clustering_biases}

Let ${N}(\phi,\,m)$ be the differential number count of a population of extragalactic sources in the magnitude range [$m$, $m+dm$] and angular space [$\phi$, $\phi+d\Omega$]. If magnitude estimates are biased with a shift $\delta m(\phi)$, the apparent number count reads
\begin{equation}
    {\hat{N}_{\rm {dered}}}(m) = {\rm N}(m+ \delta m )\;.
    \label{eq:number_counts_a}
\end{equation}
Considering a power-law luminosity function with a slope 
\begin{equation}
  \alpha(m)=2.5\, \frac{ \textrm{d\,log}\, {N}(m)}{{\rm d}m}\,,
    \label{eq:alpha}
\end{equation}
and expressing the magnitude shift in terms of a flux ratio $\mu (\phi)$ such that 
\begin{equation}
\mu = 10^{-0.4\,\delta m} 
\approx 1-0.92\, \delta m \;,
\label{eq:mu}
\end{equation}
Equation~\ref{eq:number_counts_a} becomes 
\begin{eqnarray}
    \hat{N}_{\rm dered}(m)\, &=&\, \mu^{\alpha}\, {N}(m)  \nonumber\\
    &\approx&\, (1+\alpha\, \delta \mu)\, {N}(m)\;,
    \label{eq:number_counts}
\end{eqnarray}
where the second equality is based on $\mu = 1+\delta \mu$, with fluctuations $\delta\mu$ being small compared to unity. Note that the slope of source number counts $\alpha$ is typically of order unity. To summarize, we have
\begin{equation}
 - \delta m \approx \delta \mu \approx R_{\lambda}\, \langle \delta E_{B-V}\rangle_g 
 \approx \delta A^{\rm EG}_{\lambda} \;.
\end{equation}
We point out that this formalism describing changes in number counts under brightness changes is similar to that used in gravitational-lensing magnification. Both extinction overcorrection and gravitational magnification introduce biases in number counts, density field, and angular clustering measurements for magnitude-limited samples. Considering source overdensity fluctuations over the sky, 
\begin{equation}
    \delta(\phi) = {N}(\phi)/\langle {N}\rangle -1\; ,
    \label{eq:overdensity_def}
\end{equation} Equation~\ref{eq:number_counts} shows that a biased dereddening estimate will introduce modulations in the apparent number counts. The estimated overdensity is given by
\begin{eqnarray}
    \hat{\delta}_{{\rm dered}}(\phi) \, &=& \, \delta(\phi)[1+\alpha\, \delta \mu(\phi)]+\alpha\, \delta \mu(\phi)\nonumber\\
    \,&\simeq&\, \delta(\phi)+\alpha\, \delta \mu(\phi)\,,
    \label{eq:apparent_delta_g}
\end{eqnarray}
where the $\delta \mu$ modulation could be appreciable when the density contrast $\delta$ is small. Any magnitude-limited sample of dereddened sources will thus carry spatial fluctuations of the extragalactic imprints present in the dust map used for extinction correction. As a direct consequence, angular correlations between two magnitude-limited, dereddened populations will also be affected. Let us consider the density fields $\delta_1$ and $\delta_2$ for these two populations with an intrinsic angular clustering $w_{12}= \langle \delta_{1} \cdot \delta_{2}\rangle$. Following Equation~\ref{eq:apparent_delta_g}, the bias field in the dust map will modulate the apparent two-point function such that 
\begin{eqnarray}
    \hat{w}_{12, \rm dered} \, &=&\,  \langle \hat{\delta}_{1, \rm dered} \cdot \hat{\delta}_{2, \rm dered}\rangle  \nonumber\\
    &\simeq&\, w_{12} + \alpha_2\, \langle \delta_1 \cdot \delta\mu \rangle  + \alpha_1\, \langle \delta_2 \cdot \delta\mu \rangle \nonumber\\ &+& \alpha_1\,\alpha_2\, \langle \delta\mu^2 \rangle\;,
    \label{eq:2pt_fn_bias}
\end{eqnarray}
where the second and third terms are the source--dust map bias correlations (one for each population), and the last term is the autocorrelation of the extragalactic bias in a dust map. The presence of extragalactic contamination thus sets a floor affecting two-point function measurements using magnitude-limited, dereddened extragalactic sources.

\subsection{Measuring Extragalatic Imprints in a Dust Map}

We now consider certain properties of dust maps relevant to our analysis.
\begin{enumerate}
\item Dust maps are typically provided using reddening units; hereafter, we will follow this convention and express dust column density estimations in units of $E_{B-V}$. 
\item If the zero point of the reddening is properly calibrated in a map, the spatial average of the extragalactic imprints (e.g., the cosmic IR background (CIB) monopole in IR-based maps) should already be removed. The information that can be extracted is therefore in the fluctuations of the reddening field,
\begin{eqnarray}
    \delta E_{B-V}(\phi) = \delta E_{B-V}^G(\phi)+ \delta E_{B-V}^{EG}(\phi)\nonumber \\ \equiv E_{B-V}(\phi) - \langle E_{B-V} (\phi)\rangle,
    \label{eq:delta_ebv}
\end{eqnarray}
where $\langle E_{B-V}\rangle$ is the sky average of the reddening field. As our goal is to extract extragalactic signals, $\delta E_{B-V}^{EG}$, we choose to estimate $\langle E_{B-V} \rangle$ using a running mean with a radius of $1^{\circ}$ to suppress Galactic contributions.

\item The angular autocorrelation of dust maps $\langle (\delta E_{B-V})^2 (\theta) \rangle$ is often of limited interest, as it mixes both Galactic and extragalactic fluctuations. This autocorrelation can only be used as an upper limit to the level of extragalactic contamination.

\item It is possible to separate the extragalactic component from the Galactic reddening by considering angular cross-correlations between the dust map and an external set of reference galaxies. Let $\delta_r (\phi, z)$ be the fractional overdensity field of these reference galaxies. Since the spatial distribution of dust in the Milky Way Galaxy is not expected to correlate with such a population, the reference--dust-reddening correlation reads
\begin{eqnarray}
    \langle \delta_r(\phi, z) \cdot \delta E_{B-V}(\phi+\theta) \rangle\, &=&\, 
    \langle \delta E_{B-V}^{\rm EG}(\theta, z) \rangle_r\, , 
    \label{eq:cross-correlation}
\end{eqnarray}
where the subscript $r$ refers to the positions of the reference objects. This excess-reddening estimator thus allows us to extract extragalactic imprints as a function of redshift in each of the dust maps considered. To enhance the signal-to-noise ratio of this estimator, we also construct a variance map for each $\delta E_{B-V}$ field and replace the mean with an inverse variance-weighted mean for the $\langle...\rangle$ operator (see Appendix A for details).

\end{enumerate}

\section{Data}\label{sec:data}

\subsection{Galactic Reddening Maps}\label{sec:maps}

\begin{figure*}[p]
    \begin{center}
         \includegraphics[width=0.94\textwidth]{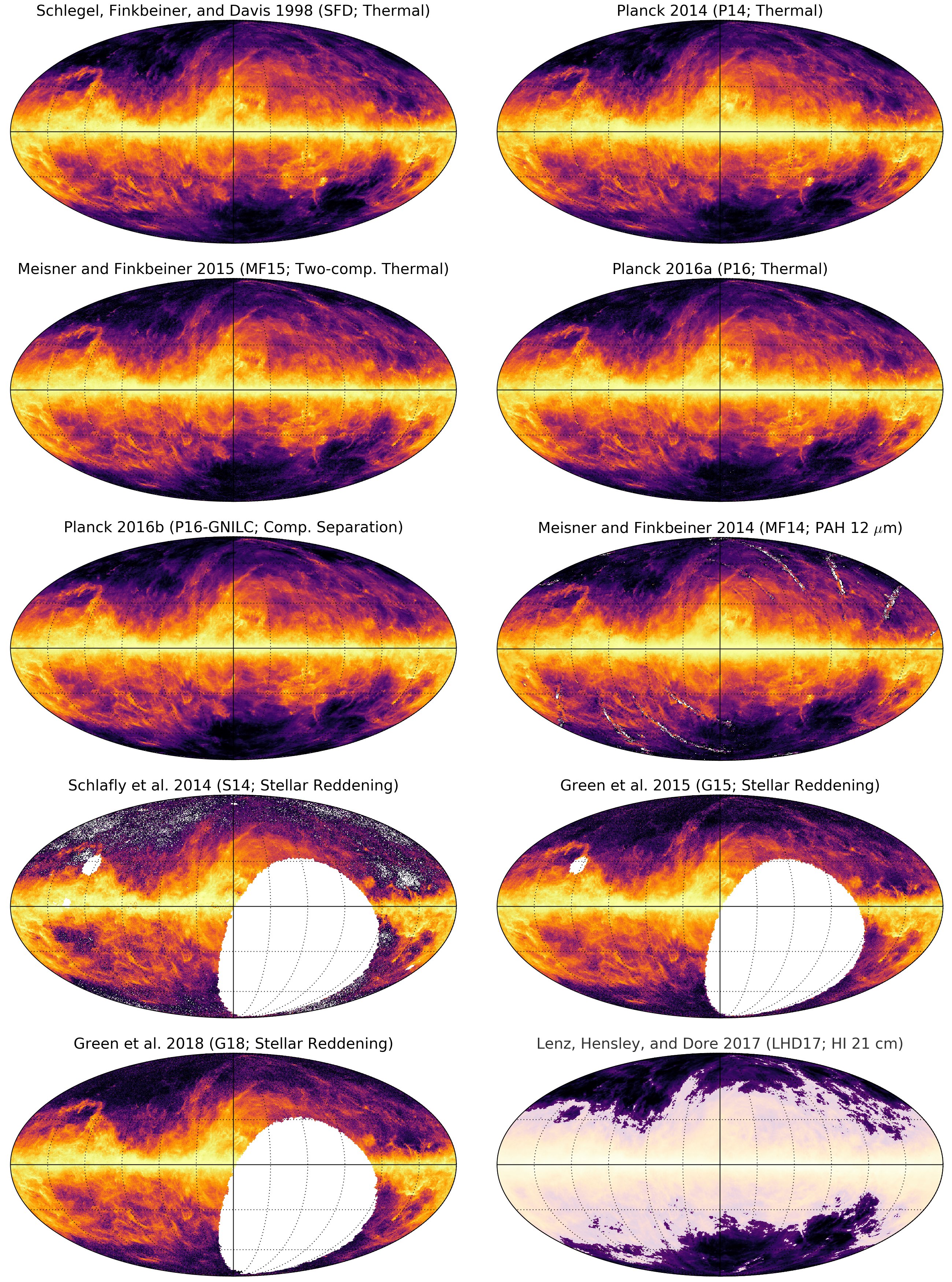}
    \end{center}
    \caption{Full-sky Mollweide projections in Galactic coordinates for the 10 Galactic reddening $E_{B-V}$ maps considered. A histogram equalization stretch is used. White patches show areas with no data or pixels with negative values due to noise. The semitransparent mask in the HI dust map indicates a high column density area where the $N_{\rm HI}$--$E_{B-V}$ becomes appreciably nonlinear.
         }
    \label{fig:maps_fullsky}
\end{figure*}

We consider the Galactic reddening maps or proxies for dust maps available in the literature and select those having data for a significant fraction of the sky and an angular resolution higher than $1^{\circ}$. This leads to a collection of 10 maps which are shown in Figure~\ref{fig:maps_fullsky} using a Mollweide projection in Galactic coordinates. Each map shows the estimated $E_{B-V}$ reddening values. These reddening maps fall into three categories: IR emission, stellar optical reddening, and HI 21cm emission. We now describe them in more detail.

\subsubsection{Dust IR Thermal Continuum Emission}\label{sec:IRmaps}

A significant fraction of the photons emitted by stars end up being absorbed by dust grains and reemitted at IR wavelengths with a modified blackbody spectrum. For a typical dust temperature of $20\,$K in the diffuse ISM, its spectral energy distribution (SED) peaks at about $150~\mu$m (see Figure~\ref{fig:dust_observables}). Since the optical depth is small in the IR and the radiation is mostly optically thin, this probe is effective over a wide range of Galactic latitudes. The conversion from dust IR emission to optical reddening requires a temperature-dependent emissivity correction, as well as a calibration to measured extinction values. Our selected IR-based dust-reddening maps are as follows.

\begin{enumerate}
\item The SFD map. This seminal work uses the full-sky 100 $\mu$m map from $IRAS$/ISSA with a $6'.1$ angular resolution. Point sources and zodiacal light are first removed from the map. An intermediate product of the dust column density map is constructed by applying a $1^{\circ}$ resolution temperature correction derived from $COBE$/DIRBE 100 and 240 $\mu$m maps. It is then converted into a reddening map calibrated using optical color and magnesium-line measurements of a sample of a few hundred elliptical galaxies. One important discovery in preprocessing the IR intensity maps is a significant detection of the CIB at 240 $\mu$m. This is revealed by the residual IR emission when extrapolating the dust--versus--Galactic HI column density \citep[as measured in the Leiden-Dwingeloo 21 cm survey in][]{1997agnh.book.....H} relation to the zero HI column.

\item The \citet[][hereafter P14]{2014A&A...571A..11P} map. It is based on a set of intensity maps with spectral sampling extended to the long-wavelength side of the dust-thermal continuum. By adding the 353, 545, and 857 GHz maps from the $Planck$ 2013 data release to the $IRAS$ 100 $\mu$m map, P14 fit a modified blackbody model and used the spectrally integrated dust radiance to derive a full-sky reddening map. Compared to SFD, P14 better captured the temperature variations of dust in different parts of the ISM via a longer spectral leverage. This could potentially lead to a smaller variance in estimating $E_{B-V}$.

\item The \citet[][hereafter MF15]{2015ApJ...798...88M} map. By using data from $IRAS$, $COBE$, and $Planck$, these authors demonstrated that a two-component model is a better description of the dust thermal emission over 100--3000 GHz (3 mm--100 $\mu$m) compared to a single modified blackbody spectrum. As a side product, MF15 derived a $6'.1$ resolution reddening map via fitting a less flexible version of the two-component model (with some parameters fixed using the global best-fit values) to the $Planck$ 217--857 GHz (2013 release) and $IRAS$ 100 $\mu$m data.

\item The \citet[][hereafter P16]{2016A&A...586A.132P} map. It is based on a physical dust model from \cite{2007ApJ...657..810D} fitted to the $WISE$ 12~$\mu$m, $IRAS$ 60 and 100~$\mu$m, and the full-mission $Planck$ maps in 857, 545, and 353~GHz. Since the model SED at mid-IR is determined by parameters almost independent of dust column, the spectral sampling for reddening constraints is essentially the same as that for P14 and MF15. The best-fit physical parameters in P16 directly determine the model extinction in each pixel, but comparisons to external extinction measurements indicate some limitations in this approach. Then, P16 renormalizes the model extinction scale to match that of the empirical measurements.

\item The \citet[][hereafter P16-GNILC]{2016A&A...596A.109P} map. This work attempts to minimize the contribution of the CIB in the inferred dust map. At high Galactic latitudes, the contribution from the CIB to the small-scale fluctuations of the IR sky is significant compared to that of the Galactic dust \citep{2011A&A...536A..18P,2014A&A...571A..30P}. Motivated by this property, this work uses a spatial prior to better separate these two components with otherwise similar SEDs. Using a component separation method called generalized needlet internal linear combination, \text{P16-GNILC} filters out small-scale structures in all of the full-mission channel maps in $Planck$, as well as the $IRAS$ 100~$\mu$m map. 
The angular resolution is adaptively reduced from the original $Planck$ beam. At high latitudes, where maximum smoothing occurred due to fractionally higher CIB, the angular resolution is $\sim 15'$. A full-sky Galactic dust opacity at 353 GHz is derived by fitting a modified blackbody model to each pixel at 353, 545, and 857 GHz and 100 $\mu$m.

\end{enumerate}
 
\subsubsection{Mid-IR PAH Emission}\label{sec:PAHmaps}

The PAH molecules are responsible for a series of emission features in $\sim3$--$20$~$\mu$m \citep{1984A&A...137L...5L}, as shown in Figure 1. Being smaller than dust grains, the emission of these molecules is more sensitive to the heating of the interstellar radiation field. They are not expected to correlate perfectly to dust column density but can be used as an indirect tracer of dust reddening. The PAH map considered here is as follows.\\

\begin{enumerate}

\item The \citet[][hereafter MF14]{2014ApJ...781....5M} map. This work aims at providing a dust map with a higher angular resolution than those based on IR emission. It uses the full-sky $WISE$ 12~$\mu$m (W3 band) data for which point sources have been removed. To alleviate the contamination from the Moon and zodiacal light, these authors recalibrated the large-scale zero point ($15^{\circ}$) in W3 using the $Planck$ 857 GHz map. The final MF14 map has a $15''$ resolution. To ease the comparison with all of the other arcminute-resolution maps, we resample MF14 to a $3.4'$ resolution.

\end{enumerate}

\subsubsection{Optical Stellar Reddening}\label{sec:Stellarmaps}
Interstellar reddening can be directly measured using background objects for which the intrinsic colors are known. Using stars that are embedded in the dusty ISM with a range of distances, one can further probe the 3D structure of the Galactic reddening field. \cite{2014ApJ...783..114G} demonstrated the feasibility of constructing such a wide-field 3D stellar reddening map using only broadband photometry for a large sample of photometrically selected stars in the Pan-STARRS1 survey \citep[PS1;][]{2016arXiv161205560C}. A probabilistic framework is introduced to model the type of a star and the distance and reddening to it simultaneously. This leads to three PS1 stellar reddening maps.

\begin{enumerate}

\item The \citet[][hereafter S14]{2014ApJ...789...15S} map. Using PS1 $grizy$ photometry for $\sim 500$ million point sources, S14 modeled the interstellar reddening integrated to a 2D plane of 4.5 kpc. At high latitudes, this distance is sufficient to incorporate nearly the full Milky Way dust column. An adaptive pixelization is used, resulting in an angular resolution of $7'$--$14'$ (mostly $14'$ at high latitudes). 

\item The \citet[][hereafter G15]{2015ApJ...810...25G} map. This work improves the modeling used in S14 and supplements the PS1 optical photometry with 2MASS near-IR $JHK_s$ bands to map interstellar reddening using $\sim 800$ million stars. A full 3D map is provided at a similar angular resolution as S14. For the present work, we integrate G15 to infinity for the full Galactic dust column (as shown in Figure 2). Compared to S14, the noise at high latitudes appears to be largely reduced.

\item The \citet[][hereafter G18]{2018MNRAS.478..651G} map. This work updates the G15 map by adding another 1.5 yr of PS1 data. An different extinction vector from \cite{1989ApJ...345..245C} and \cite{2011ApJ...737..103S} is adopted instead of that from \cite{1999PASP..111...63F} in G15, resulting in better fits to the photometry. As with G15, we integrate the 3D reddening map of G18  to the longest-distance bins to obtain a 2D reddening field of the Milky Way.

\end{enumerate}

\subsubsection{HI 21 cm Emission}\label{sec:HImaps} For diffuse ISM at high latitudes where dust and neutral gas are well-mixed, one can use the HI hyperfine 21 cm emission as an indirect dust tracer. Compared to the broad dust continuum in emission and absorption, the line nature of the HI emission provides a probe of the velocity structure of the ISM. Meanwhile, extragalactic contamination should be readily removed by applying a velocity cut to the HI data. Our selected HI-based reddening map is as follows.

\begin{enumerate}

\item The \citet[][hereafter LHD17]{2017ApJ...846...38L} map. This work derives a reddening map using the full-sky HI 21 cm map from the HI4PI survey \citep{2016A&A...594A.116H}, which merges the data from the Effelsberg Bonn HI Survey \citep[EBHIS;][]{2011AN....332..637K} and the Parkes Galactic All-Sky Survey \citep[GASS;][]{2009ApJS..181..398M}. To infer dust reddening, LHD17 investigated the scaling relation between the HI column $N_{\rm HI}$ and the $E_{B-V}$ in SFD. A linear conversion is derived that is valid for low column density regions with $N_{\rm HI} < 4 \times 10^{20}$~cm$^{-2}$ (or $E_{B-V}< 45$~mmag). These authors showed that the scatter in this relation is minimized after applying a velocity cut of 90~km~s$^{-1}$, which excludes high-velocity clouds that appear to be deficient of dust. This map has a resolution of $16'.1$. 
\end{enumerate}

\begin{figure}[t]
    \begin{center}
         \includegraphics[width=0.47\textwidth]{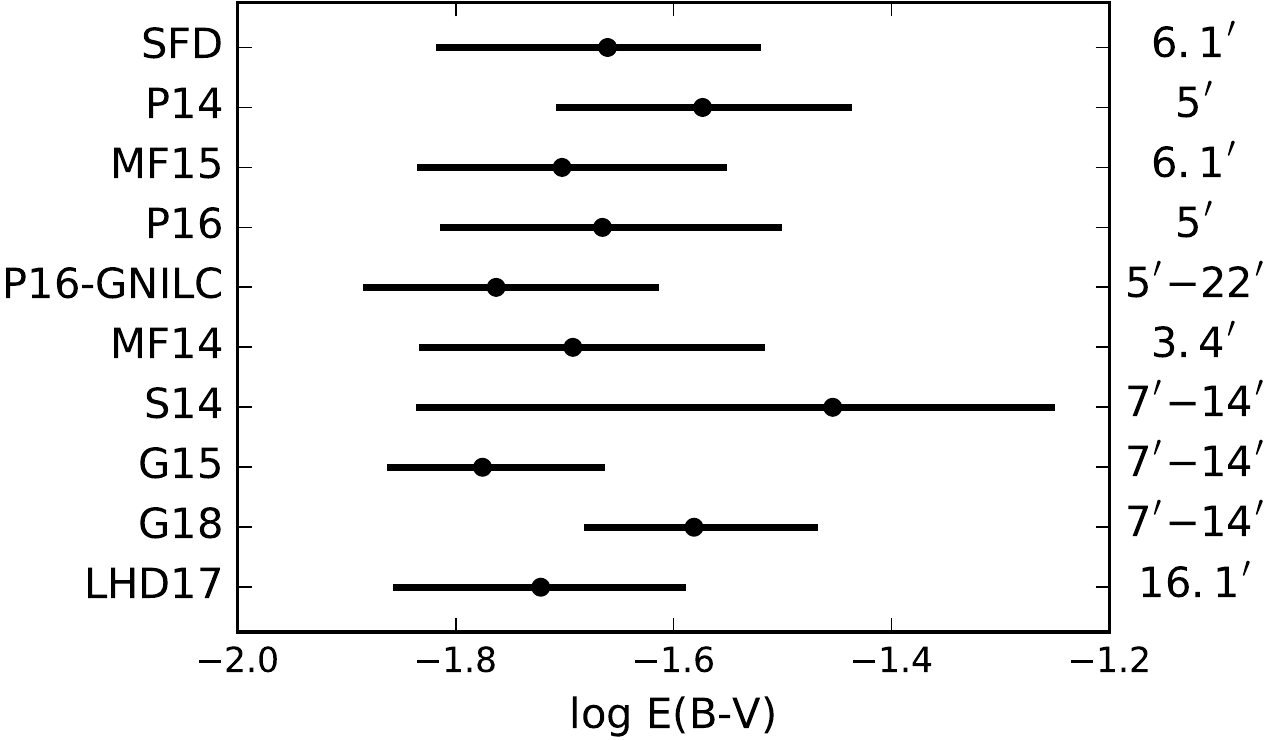}
    \end{center}
    \caption{Median and interquartile range of the $E_{B-V}$ values over the NGC area for each dust map, with the angular resolution labeled on the right $y$-axis.}
    \label{fig:pdf-res}
\end{figure}

\subsection{Processing of the Maps}\label{sec:processing}

\begin{figure*}[p]
    \centering 
         \includegraphics[width=0.96\textwidth]{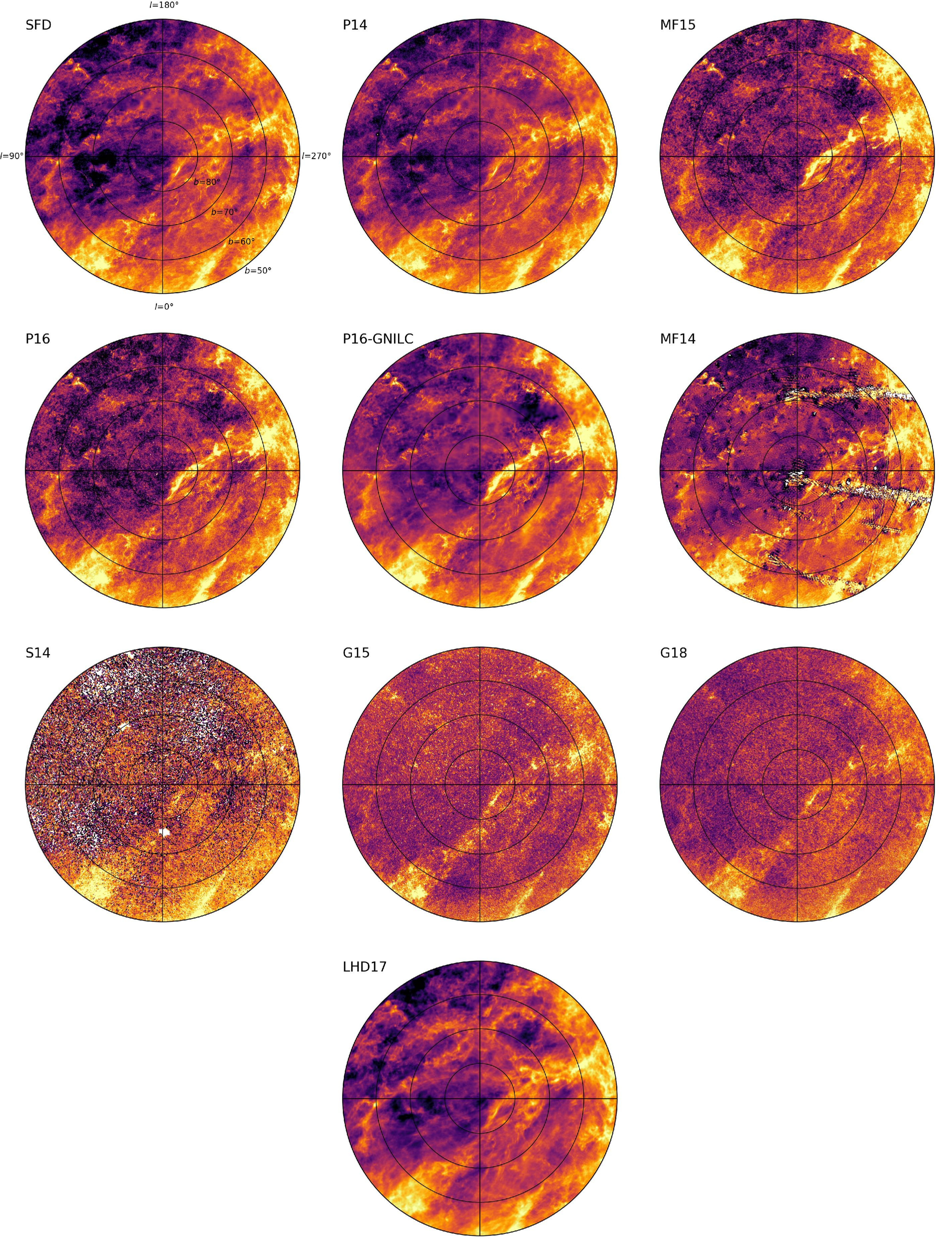}
         \caption{Lambert equal-area projections of the NGC ($b>50^{\circ}$) for the 10 Galactic reddening $E_{B-V}$ maps considered. White pixels have negative $E_{B-V}$ values due to noise.}
         \label{fig:dust_map_ngc}
\end{figure*}

To facilitate the comparison between the above maps, we post-process each product to a standard format. For the spatial sampling, we use the HEALPix scheme \citep{2005ApJ...622..759G}. We keep the intrinsic resolution for all the maps as shown in the right $y$-axis in Figure \ref{fig:pdf-res}) except for MF14 whose resolution ($15''$) has been downgraded using pixels of $3.4'$ ($N_{\rm side}=1024$). We then oversample all the maps to a common grid of $0.85'$ ($N_{\rm side}=4096$), which allows us to reduce pixelization effects when measuring small-scale cross-correlations. We consider only the north Galactic cap (NGC) area with $b>50^{\circ}$ for all maps. We convert all the maps to the $E_{B-V}$ reddening unit. For optical-depth maps, we use the conversion suggested for each product. For the MF14 map, no direct reddening calibration is provided; we convert its original 12 $\mu m$ intensity in units of MJy to reddening with a linear conversion factor such that its mean reddening over the NGC matches that in SFD. After homogenizing the map unit, the overall reddening scales and the  distributions of the $E_{B-V}$ values in the NGC in these maps are not entirely consistent due to different calibrations (Figure \ref{fig:pdf-res}). We then renormalize each map to have the same median as that in SFD over the NGC (0.0218 mag). Figure \ref{fig:dust_map_ngc} shows these renormalized maps using an area-preserving Lambert projection over the NGC area. We note that \cite{2011ApJ...737..103S} found that the $E_{B-V}$ in SFD is systematically overestimated by about 13$\%$; this correction will be considered in the discussion section but has not been applied in our reprocessed maps or in the results in Section~\ref{sec:exgalresults}.

\begin{figure}[t]
    \begin{center}
         \includegraphics[width=0.47\textwidth]{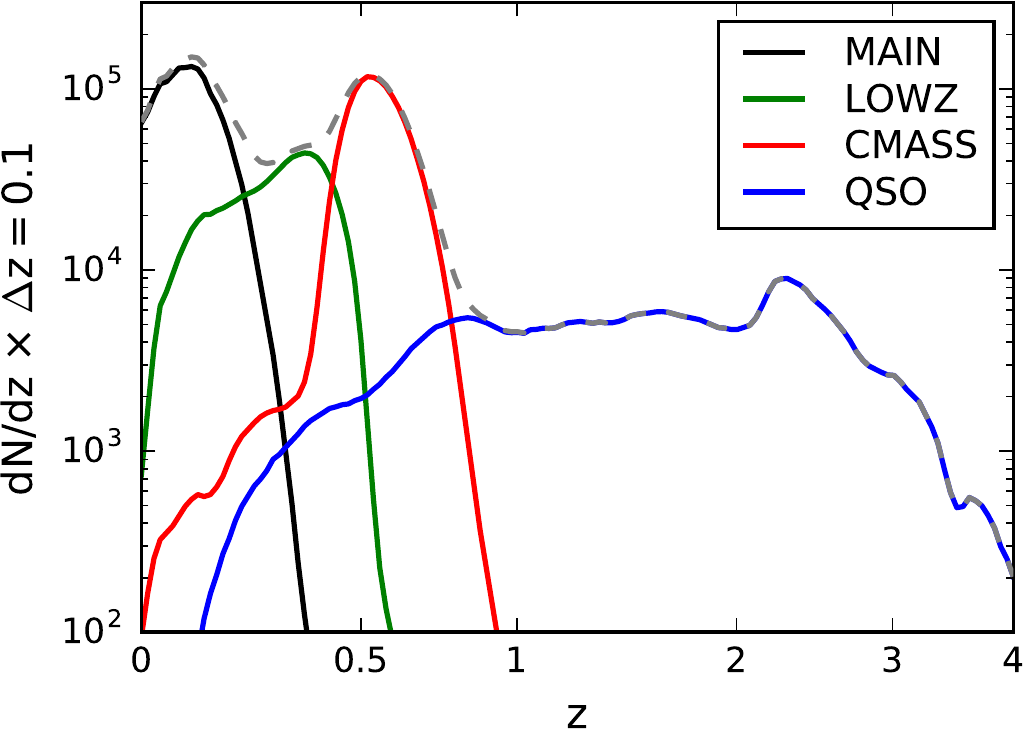}
    \end{center}
    \caption{Redshift distribution for the spectroscopic reference samples used as 3D matter tracers: SDSS MAIN galaxy sample (black), BOSS LOWZ and CMASS LRGs (green and red, respectively), and SDSS I--IV quasars (blue). The gray dashed line shows the combined sample.}
    \label{fig:reference}
\end{figure}

\subsection{Extragalactic Reference Objects}\label{sec:ref}
The goal of our analysis is to extract potential extragalactic signatures in Galactic dust maps using angular cross-correlations with tracers of the large-scale structure as a function of redshift. To select these tracers, we combine four spectroscopic samples of galaxies and quasars derived from the SDSS: the ``MAIN'' sample galaxies from the NYU value-added large-scale structure catalog from \cite{2005AJ....129.2562B} \citep[similar to that in][]{2002AJ....124.1810S}; the ``LOWZ'' and ``CMASS'' luminous red galaxy large-scale structure catalogs from \cite{2016MNRAS.455.1553R}; and ``QSO'', the DR14 quasar catalog from \cite{2018A&A...613A..51P}, which includes all the SDSS I--III quasars and new objects obtained in the ongoing SDSS IV eBOSS program. The final sample spans a wide range of redshift from 0 to about 4 (Figure \ref{fig:reference}); the total sample size is 1.1 million within the NGC. In Appendix B we present the redshift- and sample-dependent clustering bias of these reference objects. In Appendix C, we further show that the potential extinction-induced selection bias already present in the reference samples cannot account for the dust map reference cross-correlation amplitudes we find in Section~\ref{sec:exgalresults}.

\section{Results}\label{sec:exgalresults}

\begin{figure}[t!]
    \begin{center}
         \includegraphics[width=0.47\textwidth]{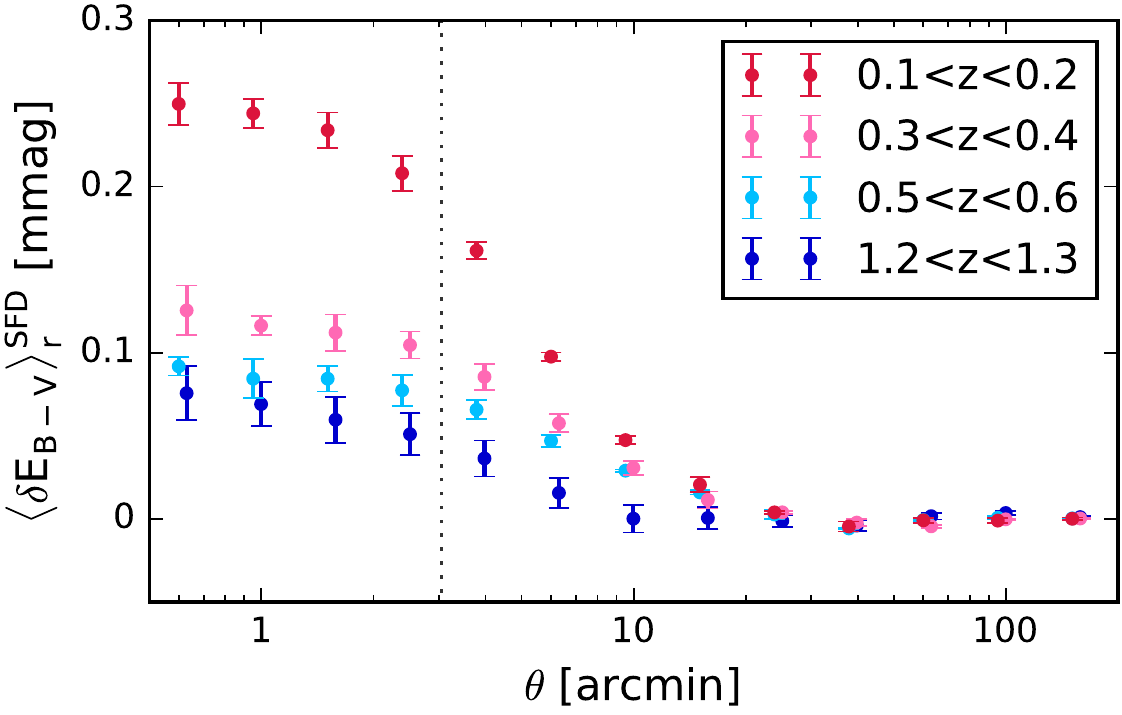}
    \end{center}
    \caption{Angular cross-correlation functions between $E_{B-V}$ in SFD and extragalactic reference objects at four redshift bins showing angular and redshift dependencies. The dotted line shows the beam size ($\rm FWHM=6'.1$) corresponding to that of the SFD map.}
    \label{fig:angular-correlation}
\end{figure}

We now present the extraction of extragalactic imprints in Galactic dust maps. For each dust map, we measure the correlation between reddening and the density of reference sources as a function of angular separation and redshift. 

\begin{enumerate}
\item Detection and angular dependence. As an example, Figure \ref{fig:angular-correlation} shows the angular cross-correlation functions we measure in the SFD map for four selected redshift bins of reference objects. In each case, a significant excess reddening is seen at scales below a few tens of arcminutes. The profile is flattened at small scales by the effective beam and at large scales ($\sim 1^{\circ}$) due to our smoothing for suppressing Galactic fluctuations. As the angular dependence can be meaningfully reconstructed by these two scales, the relevant information lies in the amplitude of the cross-correlations. To extract it, we estimate the effective excess reddening $\Delta E_{B-V}(z)$ around the reference,
\begin{equation}
    \Delta E_{B-V}(z) \equiv \frac{1}{\theta_{max}} \int_{0}^{\theta_{max}} \langle \delta E_{B-V}(\theta, z) \rangle_r \,d\theta\,,
    \label{eq:excess-reddening-redshift}
\end{equation}
where $\theta_{max}=10'$, compatible with the coarsest resolution ($\rm FWHM = 22'$) of the maps we consider. The choice of a 1D integral is motivated by the goal of extracting a representative amplitude for a dust column density or surface brightness. We thus do not perform a 2D integral used for aperture photometry. To quantify the errors, we bootstrap the reference sample and estimate the variance in the measured $\Delta E_{B-V}$. If, for a given map, $\Delta E_{B-V}$ is significantly different from zero, one can conclude that the dust map is contaminated by extragalactic signals within the corresponding redshift range.

\item Origin of the reddening excess. Over $10'$ scales ($\sim 5$ Mpc over a wide redshift range), where we choose to measure $\Delta E_{B-V}$, the signal is expected to come from hundreds of unresolved galaxies clustered around our reference objects. The amplitudes of $\Delta E_{B-V}$ are thus expected to scale with the clustering bias of the reference objects but not necessarily their star-formation rates; we show in Appendix B that this is indeed the case. The extragalactic signals we detect on Mpc scales in IR-based dust maps originate from the bulk of the CIB correlated with the reference spectroscopic objects. The corresponding IR emission is not dominated by that of the reference objects.

\end{enumerate}

\begin{figure*}[t]
    \begin{center}
         \includegraphics[width=0.68\textwidth]{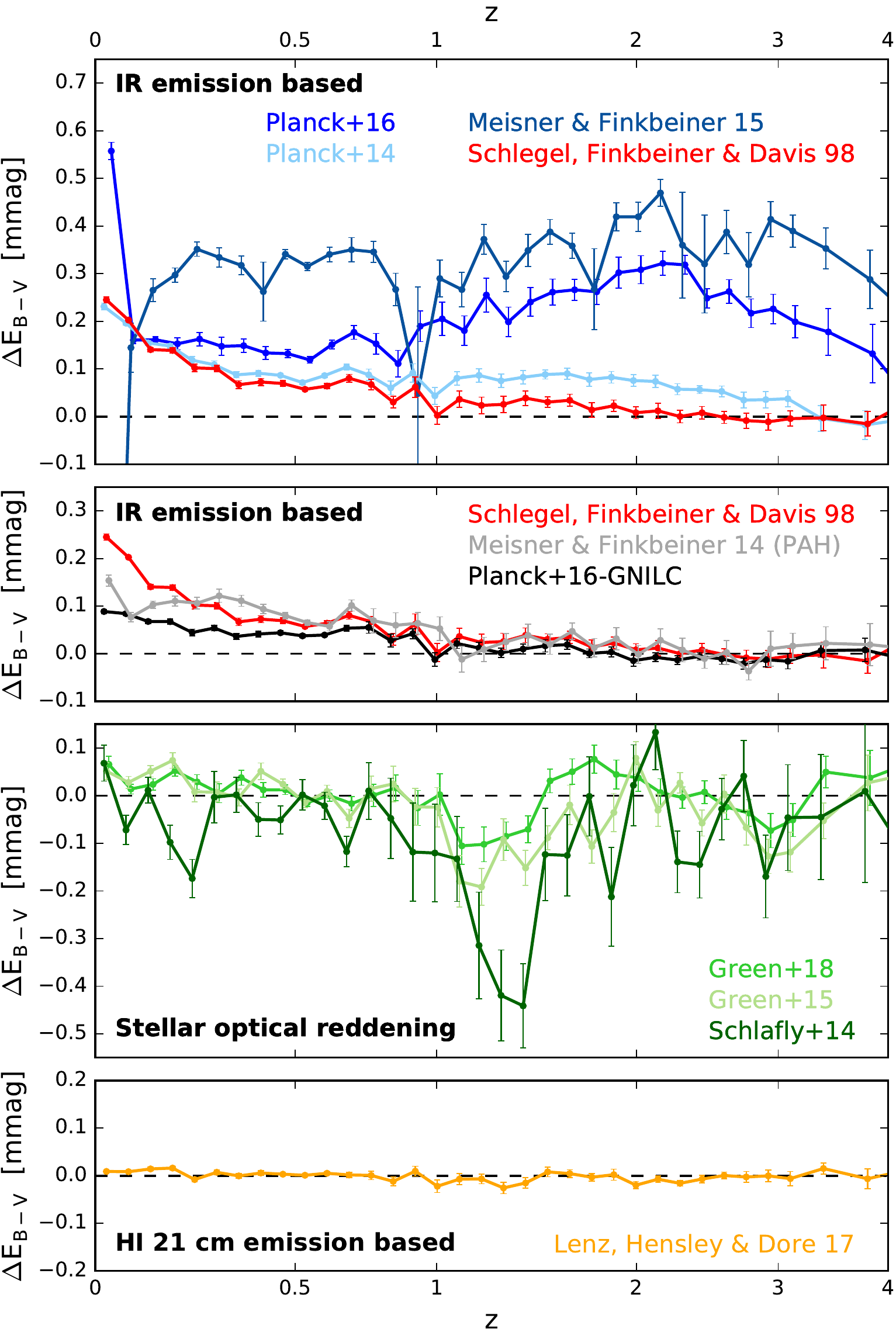}
    \end{center}
    \caption{Excess $E_{B-V}$ within $10'$ around reference objects as a function of redshift for each dust map. The results are organized into four groups. First panel: the IR dust thermal modeling maps with high extragalactic contamination. Second panel: nonstandard IR-based maps with lower extragalactic contamination, together with the classic SFD already shown in the first panel. Third panel: optical stellar reddening maps with negative correlations at $1<z<1.5$. Fourth panel: the HI-based dust map.}
    \label{fig:excess-ebv-redshift}
\end{figure*}

Below, we describe our findings for the different types of dust maps, i.e. based on IR emission, PAH emission, stellar reddening, and 21cm emission. Figure~\ref{fig:excess-ebv-redshift} presents the observational results for all of the dust maps.

\subsection{IR-based Dust Maps}\label{sec:resultsIR}

Figure~\ref{fig:excess-ebv-redshift} shows the mean reddening excess found around spectroscopic reference objects as a function of redshift for each of the 10 dust maps considered. The excess reddening found for the SFD dust map is shown in red in the first and second panels. We observe an extragalactic contribution extending up  to redshift $2$. At low redshift, the amplitude of this reddening excess is about $0.25$ mmag. Given that the mean $E_{B-V}$ over the NGC area is about 20 mmag, this excess corresponds to a CIB contamination at the few percent level on $10'$ scales around the extragalactic reference objects. The declining redshift trend of the extragalactic imprints in SFD is consistent with the effect of a ``$K$-correction'' \citep{1956AJ.....61...97H}, given the spectral sampling of 100 $\mu$m used to constrain the dust column density in SFD. As a roughly $20$K blackbody spectrum gets redshifted, the 100 $\mu$m band probes the Wien side of the CIB SED, thus receiving less flux emitted at higher redshifts.

The redshift tomography results for the other five IR thermal dust maps are shown in the first and second panels of Figure~\ref{fig:excess-ebv-redshift} together with that in SFD. The first panel compares different products using classic thermal modeling: the $IRAS$-based SFD and three recent maps incorporating additional $Planck$ data. P14 (light blue) shows an excess reddening similar to that in SFD at low redshifts, but dropping off more slowly at high redshifts. The $\Delta E_{B-V}$ in P16 (blue) appears flat at low redshifts and significantly increases and peaks at about $0.3$ mmag at $z=2$, several times higher than that in P14 at the same redshift. The two component modeling map MF15 (dark blue) shows a high and overall flat $\Delta E_{B-V}$ over the redshift range 0--4. The extreme values in the first redshift bins in MF15 and P16 appear to be representative characteristics of these map products possibly due to different treatments of bright source masking. In contrast, the dip at $z\sim0.9$ in MF15 is not significant and is driven by a small number of outlier pixels. 

Interestingly, by adding $Planck$ data, these three thermal dust maps show higher contamination at high redshift compared to SFD. This can be understood as the $Planck$ bands sample the Rayleigh-Jeans side of the dust emission SED (Figure \ref{fig:dust_observables}); extragalactic dust would thus receive a negative $K$-correction over a broad range of redshift. In addition, high-redshift $\Delta E_{B-V}$ is further boosted, since the intrinsic CIB source intensity increases largely from $z=0$ to $z\sim2$--$3$ following the cosmic star-formation history \citep{2014ARA&A..52..415M,2015MNRAS.446.2696S}. Such thermal dust maps cannot distinguish between cold Galactic dust and warm redshifted dust due to the degeneracy between temperature and redshift. Therefore, although these more recent maps using $Planck$ data better capture the reddening due to cold dust in the Milky Way, they also suffer from higher bias from the extragalactic background. The difference between the two $Planck$ team products is substantial. This is likely due to two reasons. First, P14 uses the shallower $Planck$ 2013 data release instead of the full-mission data used in P16. The $Planck$ bands thus contribute to a higher fraction of the signals in P16 when combining with the $IRAS$ 100 $\mu$m data. Second, the $E_{B-V}$ calibration in P16 effectively has a strong temperature correction for cold, low-emissivity dust to be as effective in absorption; redshifted CIB contamination thus becomes more prominent in P16. Along the same lines, MF15 further amplifies CIB contamination by capturing it with an additional ``cold'' component and interpreting all this cold dust to be of Galactic origin.

The second panel of Figure \ref{fig:excess-ebv-redshift} compares two nonstandard IR-based dust maps---P16-GNILC and MF14---with SFD. For the $Planck$ component separation dust map P16-GNILC (black), the excess reddening is detected at the level of a factor of $1.5$--$2$ lower $\Delta E_{B-V}$ than that in SFD, and becomes insignificant at $z>1$. This result suggests that indeed, this component separation algorithm is effective in filtering out CIB fluctuations; however, it is still not completely clean. The suppression of extragalactic bias is obtained at the price of losing the angular resolution of the true Galactic features, thus potentially leading to a higher variance compared to other maps. Our analysis provides a way to guide the future developments of component separation and test the performance of different algorithms.

\subsection{PAH-based Dust Map}\label{sec:resultsPAH}

Our tomographic analysis of the \cite{2014ApJ...781....5M} map (MF14) also shows an extragalactic signal up to at least $z\sim1$ (second panel in Figure~\ref{fig:excess-ebv-redshift}). We point out that a significant signal can be detected up to $z\sim2$ if we use wider redshift bins. This map is directly based on $WISE$ 12~$\mu$m observations that trace the PAH emission. Our measurement therefore provides us with a detection of the unresolved extragalactic PAH background. We note that its redshift dependence is similar to that of the SFD map. This matches expectations: due to a wide filter bandpass of the $WISE$ 12~$\mu$m channel, the flux density changes smoothly when the spiky PAH features move in and out of the band as the redshift increases. The low-redshift decline in MF14 is slightly slower than that in SFD because the slope of the overall SED leftward of 12~$\mu$m is shallower than that around 100~$\mu$m sampled by SFD (Figure 1). At redshifts below and beyond $\sim 1$, the slopes of the excess in MF14 appear different; this transition coincides with the redshift where the bluest PAH feature at 3.3~$\mu$m enters the $WISE$ 12~$\mu$m band. 
When renormalizing the amplitude of the dust maps (see Section~\ref{sec:processing}), we scaled the $WISE$ 12~$\mu$m intensity by a Galactic dust-to-PAH ratio to match the mean $E_{B-V}$ in SFD over the NGC. As the overall amplitude of the extragalactic reddening excess in MF14 is similar to that found in SFD, it shows that the PAH-to-dust fraction in the Milky Way is comparable to that of the cosmic mean.

\subsection{Stellar Optical Reddening Dust Maps}\label{sec:resultsStellar}

We show the results for the three PS1 ``stellar" reddening maps, S14, G15, and G18 in the third panel of Figure 7. Interestingly, we find extragalactic imprints in in all of these maps, while this time the effect is opposite to what was found in IR emission maps. Our analysis shows that the Galactic reddening is underestimated at certain redshifts. These ``stellar" reddening maps are based on the Pan-STARRS point-source catalog. Such a selection inevitably includes a number of unresolved galaxies and quasars in the sample. The analyses performed in S14, G15 and G18 use a probabilistic framework to infer the objects' intrinsic colors and line-of-sight reddening induced by Galactic dust. The fact that we find a negative $\Delta E_{B-V}$ at $1<z<1.5$ suggests that at these redshifts, there is a population of objects intrinsically bluer than the stellar color locus that could be scattered inward to bias the reddening estimations. The fact that all of our $1<z<1.5$ reference objects are quasars does not necessarily mean that the contaminants in the star catalogs are quasars; they could be unresolved galaxies, which might actually outnumber quasars at these redshifts and magnitude limits \citep{2012ApJ...760...15F}.

\begin{figure}[t]
    \begin{center}
         \includegraphics[width=0.48\textwidth]{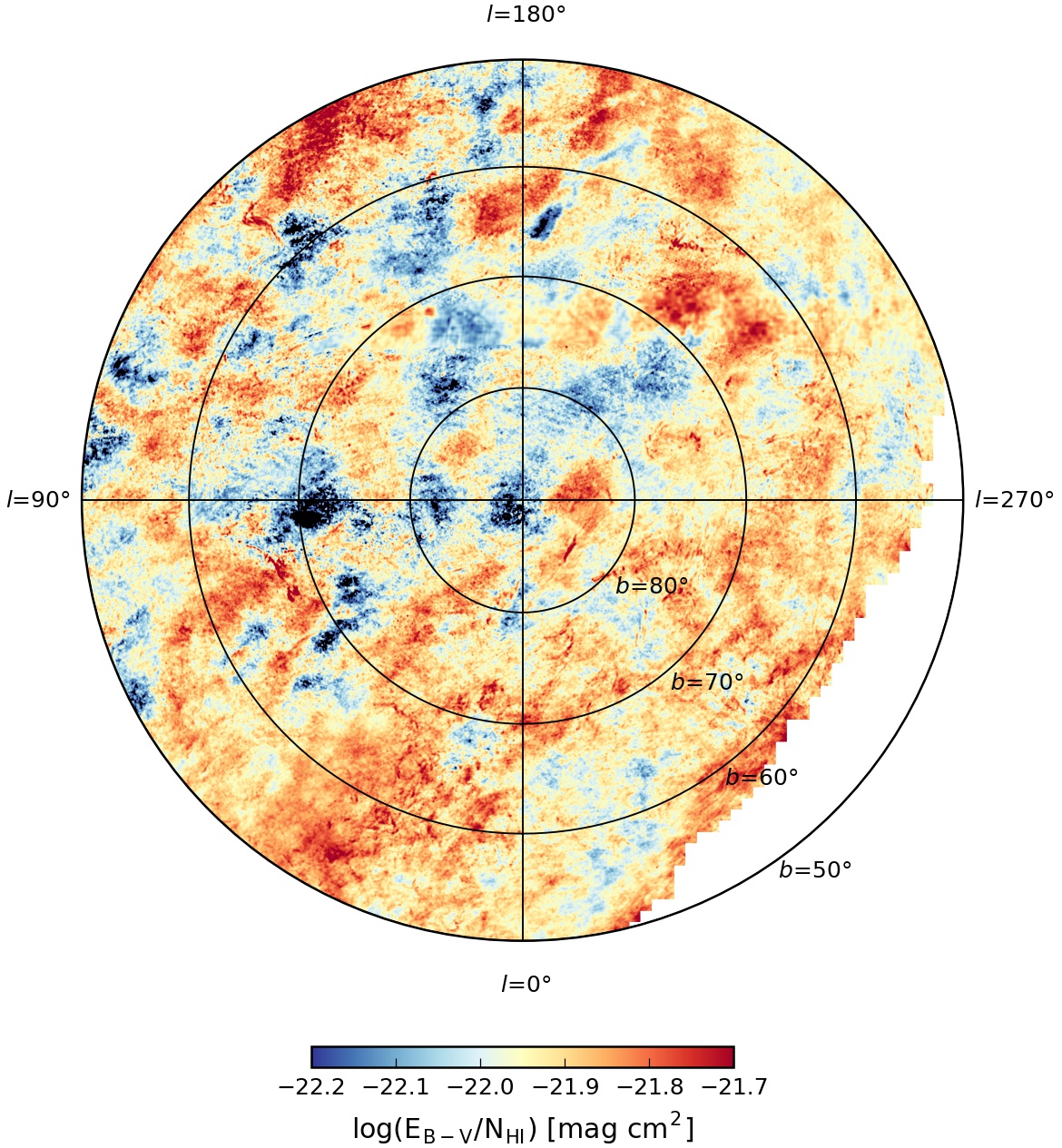}
    \end{center}
    \caption{Dust-to-gas ratio in log($E_{B-V}/N_{\rm HI}$) over the NGC area, with $E_{B-V}$ taken from SFD with the large-scale correction from \cite{2010ApJ...719..415P} and $N_{HI}$ from the HI4PI data within 90 $\rm km\,s^{-1}$.}
    \label{fig:dust_to_gas}
\end{figure}

\subsection{HI-based Dust Map}\label{sec:resultsHI}

\newcommand\Tlarge{\rule{0pt}{5.ex}}       
\newcommand\Blarge{\rule[-2.ex]{0pt}{0pt}} 

\newcommand\T{\rule{0pt}{3.8ex}}       
\newcommand\B{\rule[-2.5ex]{0pt}{0pt}} 

\begin{table*}[t!]
\caption{Extinction Overcorrection vs. Gravitational Lensing}
\centering
\label{my-label}
\begin{tabular}{ccc}

\multicolumn{1}{c}{}  & \multicolumn{1}{c}{\textbf{Extinction Overcorrection}}                                                                                            & \multicolumn{1}{c}{\textbf{Gravitational Magnification}}                                                                                                                \Tlarge\Blarge \\ \hline  
                                       
\multicolumn{1}{|c|}{Flux change ${\hat{f}}/{f}$}  &
\multicolumn{1}{c|}{\begin{tabular}[c]{@{}c@{}}$ 1 + \delta \mu$ \end{tabular}} &
\multicolumn{1}{c|}{\begin{tabular}[c]{@{}c@{}}$  1+\delta \mu$ \end{tabular}} \T\B \\ \hline

\multicolumn{1}{|c|}{Number count bias $\hat{\rm N}(m)/{\rm N}(m)$}        & \multicolumn{1}{c|}{$1+\alpha\, \delta \mu$}                                                                                          & \multicolumn{1}{c|}{$1+(\alpha-1)\delta\mu$}                                                                                                           \T\B \\ \hline
\multicolumn{1}{|c|}{Clustering bias---cross $g_f$, $g_b$ ($z_f<z_b$)}          & \multicolumn{1}{c|}{$\hat{w}_{fb} = \alpha_b\, \langle \delta \mu \rangle_{f} +  \alpha_f\, \langle \delta \mu \rangle_{b} + \alpha_f\, \alpha_b\, \langle \delta \mu^2 \rangle$}                                                                                    & \multicolumn{1}{c|}{$\hat{w}_{fb} = (\alpha_b-1)\, \langle \delta \mu \rangle_{f}$}                                                                                                           \T\B \\ \hline
\multicolumn{1}{|c|}{Clustering bias---autocorrelation}          & \multicolumn{1}{c|}{$\hat{w}_{gg} = w_{gg} + 2\, \alpha\, \langle \delta \mu \rangle_{g} + \alpha^2 \, \langle \delta \mu^2 \rangle$}                                                                                    & \multicolumn{1}{c|}{$\hat{w}_{gg} = w_{gg}$ (no effect)}                                                                                                           \T\B \\ \hline
\end{tabular}
\bigskip
\tablecomments{Comparison between extinction overcorrection and magnification induced by gravitational lensing. For the extinction overcorrection the term $\delta \mu (\lambda)$ represents the differential brightening factor and is given by $0.92\,R_{\lambda}\,\delta E_{B-V}^{\rm EG}$. For lensing, $\delta \mu$  is the achromatic gravitational magnification. The coefficients $\alpha_i$ are the slopes of the source number counts as a function of magnitude.}
\label{table:lensing}
\end{table*}

The result for the HI-based dust map LHD17 are shown in the fourth panel of Figure 7. This time, we do not see any clear signature of extragalactic contamination at the level seen for the other maps. The noise level can be translated into an upper limit for the fractional extragalactic contamination in LHD17 of about $5\times10^{-4}$ over the NGC. Since this dust-reddening map estimate is a direct linear conversion from the HI column density probed by 21 cm emission within 90~km~s$^{-1}$, we can place the same upper limit for the fractional extragalactic contamination in the HI column density map. Our result shows that, with an emission-line feature, redshifted extragalactic background can be readily separated from the Galactic structures. We note that this works well only if no other strong line is present within a factor of a few blueward of the targeted line wavelength. 

A main drawback of HI-based dust maps is the reliance on an assumed dust-to-gas ratio. LHD17 used a constant dust-to-gas ratio over the sky, which is only valid up to a certain accuracy. Figure~\ref{fig:dust_to_gas} shows the dust-to-gas ratio map over the NGC area in log($E_{B-V}/N_{\rm HI}$), where the reddening is taken form SFD with large scale ($>4.5^{\circ}$) correction from \cite{2010ApJ...719..415P} calibrated using the colors of passive galaxies; the HI column density is taken from HI4PI within 90 $\rm km\,s^{-1}$, the velocity cut used in LHD17. Over the NGC, we find fluctuations that display spatial coherence over a wide range of angular scales. Over that area, we measure a scatter of 0.08 dex in $E_{B-V}/N_{\rm HI}$. This is likely due to variations in cloud properties. Given the mean $E_{B-V}$ of about 20 mmag over the NGC, this fractional scatter corresponds to an error of about $3$ mmag in $E_{B-V}$. This is larger than the magnitude offsets due to the extragalactic contamination found in the previous nine dust maps. We note that for classic IR-based maps, the modulations of dust temperature might affect the reddening estimations in a spatially coherent way similar to that of the dust-to-gas ratio modulations in HI-based reddening maps, albeit on a slightly smaller amplitude \citep{2010ApJ...719..415P}. Therefore, in order to construct samples of extragalactic sources with the most accurate magnitude estimates, one should select sources for which the variance in dust-to-gas ratio in HI-based maps (or temperature variation in IR-based maps) is minimized. As shown in Figure~\ref{fig:dust_to_gas} the large-scale variations of the dust-to-gas ratio distribution might restrict the spatial distribution of sources that can be homogeneously corrected for Galactic dust extinction.

\section{Discussion}\label{sec:discussion}

We have shown that, out of the ten wide-field Galactic dust maps currently available, nine present detectable extragalactic contamination. The remaining one, based on the hydrogen distribution, relies on an assumed dust-to-gas ratio that can be shown to have a complex spatial distribution over the sky. What are the impacts of these extinction overcorrection biases on astronomical experiments? As presented in section~\ref{sec:number_counts_and_clustering_biases}, a reddening overcorrection leads to a whole hierarchy of number count bias, overdensity bias, and clustering bias (Equations~\ref{eq:number_counts}--\ref{eq:2pt_fn_bias}). The leading terms of these biases are all of the same order: $R_{\lambda}\, \langle \delta E_{B-V}\rangle_g \approx -\delta m \approx \alpha\,\delta \mu$. The exact amplitude of this effect depends on the chosen dust map, wavelength, redshift, source population, and angular scale considered. In many cases, they are found to be at the millimagnitude level, i.e. of order $10^{-3}$ in the optical, assuming $R_{\lambda} \approx 3$ and $\alpha$ of order unity. 

\begin{figure*}[!th]
    \includegraphics[width=0.98\textwidth]{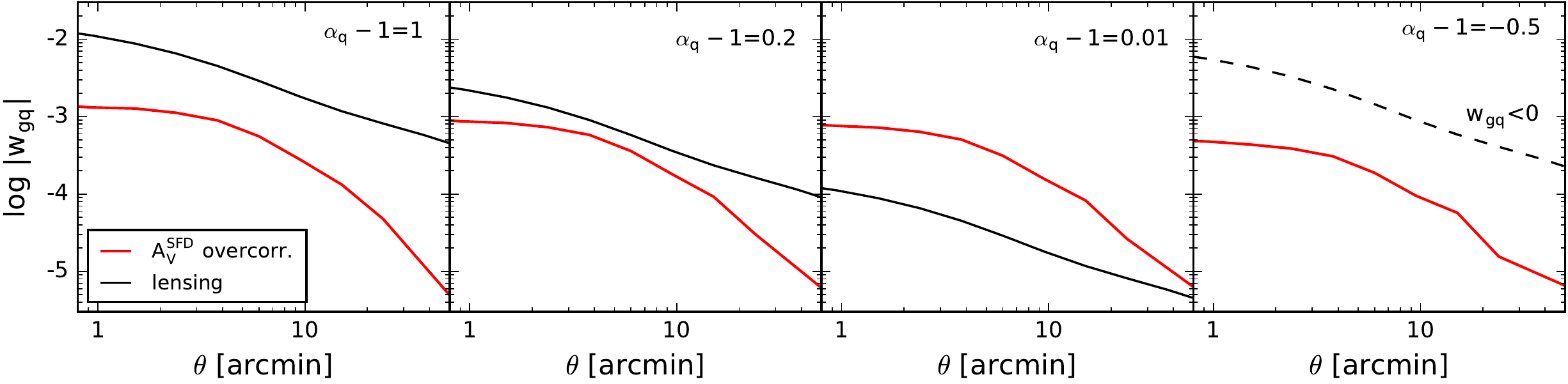}
    \caption{Case comparison for a galaxy--quasar clustering introduced by gravitational magnification (black curves; theoretical expectations) and extinction overcorrection (red curves for the $V$-band SFD bias scaled from our $\Delta E_{B-V}$ measurements). The redshifts of the galaxies and quasars are set to 0.2 and 1.5, respectively. The results are shown for a range of $\alpha_q$, the logarithmic slope of the quasar magnitude distribution, where $\alpha_q-1=0.2$ (second panel) is roughly the effective value for an optimal estimator using quasars with $g<21$.}
    \label{fig:lensing}
\end{figure*}

Systematic shifts of order millimagnitude can potentially impact precision cosmology experiments, for example, using standard candles, where the sample mean brightness has to be measured with a fractional error comparable to the targeted precision in certain cosmological parameters. Similarly, photometric offsets can affect clustering measurements such as spatial correlation functions, which are used for a wide range of applications including galaxy--halo connection, baryonic acoustic oscillations, gravitational lensing, neutrino masses, cold/warm dark matter, etc., all requiring extinction correction for the tracer sample of the experiments (various kinds of galaxies), while not all of them will be affected by a small bias. Equation~\ref{eq:apparent_delta_g} provides a rule of thumb to identify the regime in which extinction overcorrection bias will be significant: since the bias in the source overdensity $\delta (\phi)$ is roughly $\alpha\, \delta \mu(\phi)$, the bias is only important in a weak-field regime where $\delta (\phi)$ is not much greater than $\alpha\, \delta \mu(\phi)\sim 10^{-3}$ (in the optical). For a typical galaxy overdensity in 3D, $\delta (\phi)$ fluctuates at a level greater than unity on arcminute scales. However, this amplitude decreases with an increasing level of line-of-sight projection; experiments involving a substantial spread in redshift, either imposed by the photometric selection or due to photometric redshift errors, will end up falling in that category. Another relevant context is gravitational lensing-induced correlation functions, which we discuss in more detail below. Finally, we can also imagine a limiting case with a hypothetical population of randomly distributed sources (i.e., zero intrinsic clustering). When estimating the correlation function of such a population after correcting for Galactic extinction, one will end up measuring the clustering of the extragalactic contamination imprinted in the dust map.

\begin{figure*}[t!]
    \includegraphics[width=0.98\textwidth]{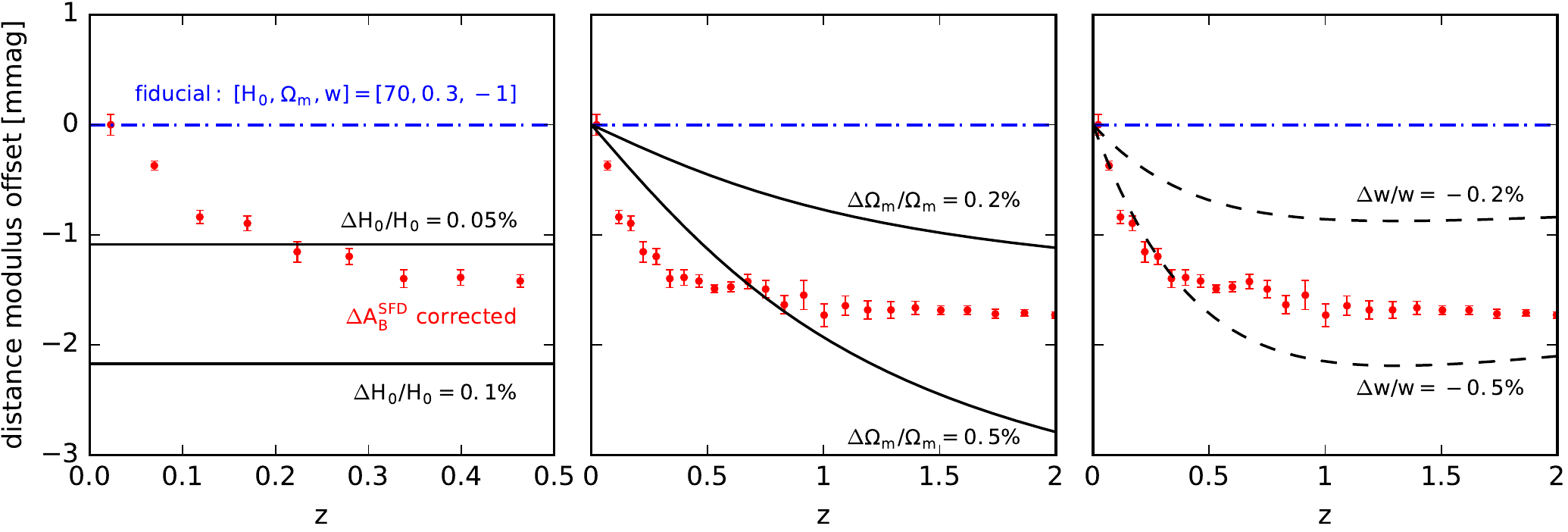}\\
    \caption{Changes in distance modulus after correcting for a rest-frame $B$-band extinction overcorrection in SFD (red data points). Black lines overlay the changes corresponding to perturbations in each of the cosmological parameters $H_0$ (left), $\Omega_m$ (middle), and $w$ (right) from a fiducial cosmology (blue dash-dotted lines). After correcting for the extinction overcorrection, one expects $H_0$ to increase by $<0.1\%$ and $\Omega_m$ and $w$ to decrease by a fraction of a percent.}
    \label{fig:sn}
\end{figure*}

\subsection{Impact on Weak-field Clustering: Lensing Magnification}\label{sec:discussion_lensing}

Gravitational-lensing magnification modulates the apparent number count of background sources by two mechanisms: magnitude brightening and solid-angle dilatation. Analogs to Equation~\ref{eq:number_counts}, this is expressed by
\begin{equation}
    {\rm N}_{\rm obs}(m)= [1+(\alpha - 1)\, \delta \mu]\, {\rm N}(m)
    \label{eq:number_counts_lensing}
\end{equation}
\citep[e.g.,][]{2005ApJ...633..589S}, where the slope of the source number counts $\alpha$ is defined in Equation~\ref{eq:alpha}, the $-1$ term takes into account the source dilution due to area dilatation, and $\delta \mu$ is the lensing magnification factor. This is similar to the formalism introduced in Section~\ref{sec:formalism} for the extinction overcorrection. The only difference is the change in sky solid angle.
The magnification effect leads to apparent angular correlations between foreground lenses and background sources that are physically uncorrelated. This can be expressed as a special case of Equation~\ref{eq:2pt_fn_bias},
\begin{eqnarray}
    w_{fb, \rm obs}(\theta) \, &=&\,  \langle \delta_{f}(\phi) \cdot \delta_{b, \rm obs}(\phi+\theta)\rangle \\
    &=&\, (\alpha_b-1)\, \langle \delta_f (\phi) \cdot \delta \mu (\phi+\theta)\rangle \nonumber\\ &=&\, (\alpha_b-1)\, b_f\, w_{\mu m}(\theta)\,,
    \label{eq:2pt_fn_bias-lensing}
\end{eqnarray}
where only the background overdensity is modulated, and the third equality assumes a linear bias $b_f$ relating $\delta_f$ to matter overdensity $\delta_m$. The gravitational potential associated with the foreground lenses introduces a magnification field $w_{\mu m}(\theta)$ to the sources, whose expression can be found in, e.g., \cite{2001PhR...340..291B}. Depending on the value of $\alpha_b - 1$, lensing-induced clustering can be positive, zero, or negative. Table~\ref{table:lensing} summarizes the effects of extinction overcorrection and lensing magnification in parallel, and one can see the similarities between the two. Unlike lensing, however, extinction overcorrection is chromatic. Another difference between the two is that the lensing efficiency is a function of the angular diameter distances of both populations, while the line-of-sight efficiency of extinction overcorrection is a constant as one applies such a correction using 2D dust maps. Given this constant efficiency, extinction overcorrection also affects source autocorrelations.

The extinction overcorrection will affect measurements of lensing-induced correlation functions. To estimate the level at which these effects occur, we examine the following scenario commonly targeted in lensing studies (see, for example, \citealt{2005ApJ...633..589S}). We consider a population of foreground galaxies at $z=0.2$ and background quasars at $z=1.5$. The magnification-induced angular correlation function expected between these two populations is shown in Figure~\ref{fig:lensing} (black curves) for several magnitude bins that correspond to several values of $\alpha_q-1$ following Equation~\ref{eq:2pt_fn_bias-lensing}. Again, $\alpha_q$ is given by the shape of the quasar luminosity function following Equation~\ref{eq:alpha}, and the range of $\alpha_q-1$ in Figure~\ref{fig:lensing} corresponds to that for quasars from $g$ of 17--21 mag going from the left to right panels. Red curves show the expected $w_{gq}$ introduced by extinction or dereddening overcorrection in the $V$ band if the SFD map is used. This is based on our excess-reddening measurements presented in Section~\ref{sec:exgalresults} assuming an $R_V$ of 3.1. As provided in Equation~\ref{eq:2pt_fn_bias}, the extinction correction--induced correlation has two terms: one that scales with $\alpha_q$ as in lensing and an additional $\alpha_g$ term. Here we set $\alpha_g$ to 1 for all panels. For large $\alpha_q$ (left panels), lensing dominates over extinction overcorrection. When $\alpha_q$ approaches 1, however, lensing effects vanish, while the extinction overcorrection is only slightly reduced, since the $\alpha_g$ term has not changed. The extinction overcorrection can thus bias lensing magnification measurements, especially at the faint end (right panels). For an optimal lensing estimator weighted by the number of quasars and the expected signal using quasars brighter than $g$ of $21$ (SDSS depth), the effective $\alpha_q-1$ is about 0.2 \citep{2002A&A...386..784M,2005ApJ...633..589S}. This is shown in the second panel, where the extinction correction--induced correlation is about $50\%$ of that induced by lensing. We thus already expect some impacts biasing current lensing measurements. Upcoming surveys such as LSST \citep{2009arXiv0912.0201L}, $Euclid$ \citep{2011arXiv1110.3193L} and $WFIRST$ \citep{2013arXiv1305.5422S} will provide us with photometric samples enabling lensing-induced correlation measurements with a precision largely surpassing that of existing measurements. In this regime, the extinction overcorrection discussed above will become a significant limitation in harnessing the full statistical power of the expected datasets. It will be important to correct or take this effect into account.

\begin{figure*}[t!]
    \begin{center}
         \includegraphics[width=0.68\textwidth]{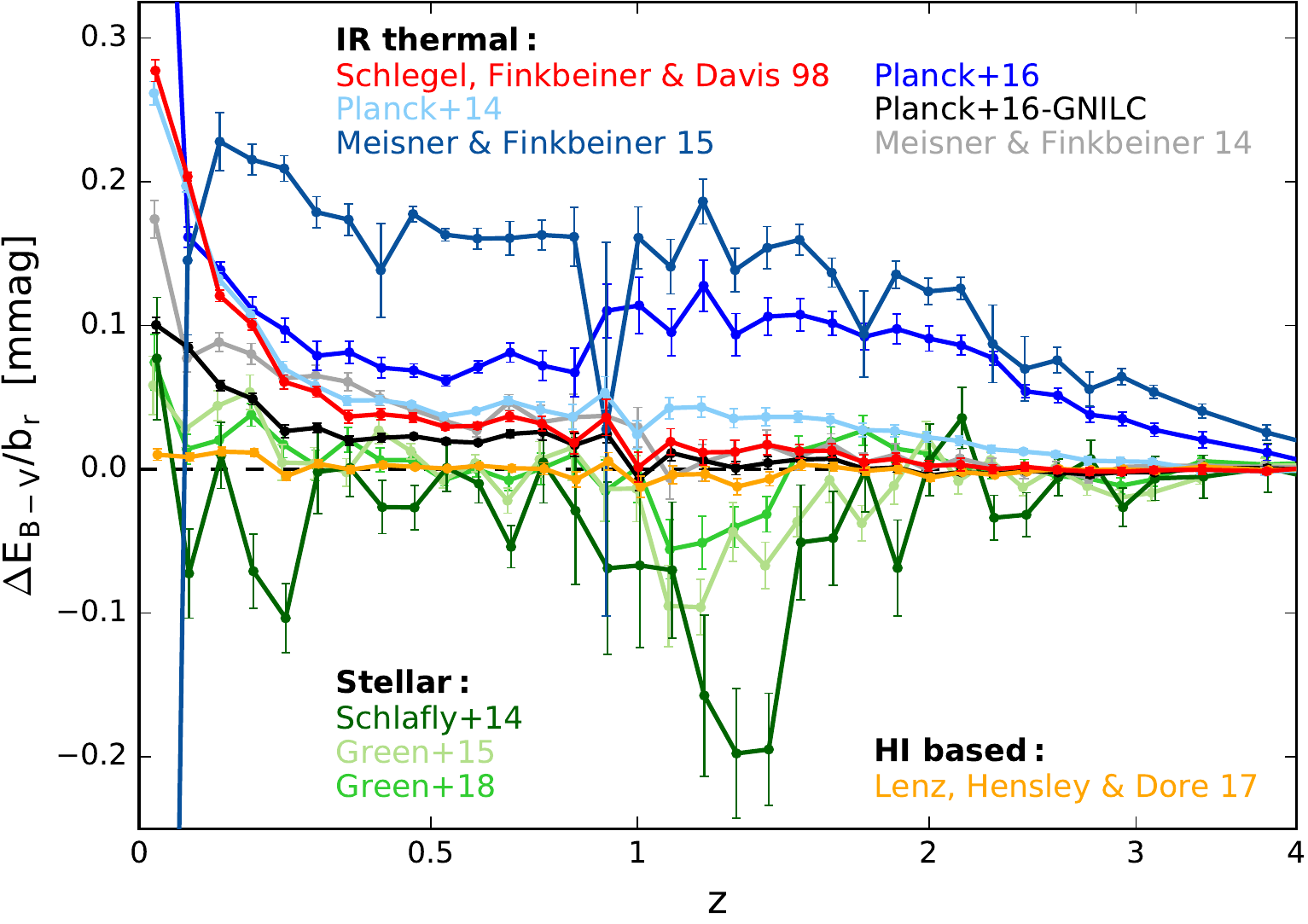}
    \end{center}
    \caption{Normalized excess $E_{B-V}$ (within $10'$) for all 10 reddening maps. The effect of our choice of reference samples has been removed by dividing out the linear galaxy bias. One can use these values for generic corrections of the dust extinction overcorrection bias present in each map.}
    \label{fig:normalized_delta_ebv}
\end{figure*}

\subsection{Cosmological Parameters with Standard Candles}\label{sec:discussionSNIa}

The extinction overcorrection is expected to impact cosmological parameter extractions using Type Ia supernovae as standard candles. Such studies utilize the cosmology dependence of the luminosity distance, thus distance modulus as a function of redshift \citep[see a review in][]{2011ARNPS..61..251G}. Briefly, the distance modulus\footnote{Unfortunately, by convention, the notation $\mu$ is the same with that for the magnification factor in the context of lensing.} $\mu$ of a standard candle with an absolute magnitude $M$ can be used to probe its luminosity distance $D_{\rm L}$:
\begin{equation}
    \mu = m - M =5\, \textrm{log} \left(\frac{D_{\rm L}(z)}{10~\textrm{pc}}\right),
\end{equation}
where $m$ is its apparent magnitude. The luminosity distance $D_{\rm L}$ as a function of redshift is cosmology-dependent,
\begin{equation}
    D_{\rm L}(z) = \frac{c}{H_0}\, (1+z)\,  \int_0^z\frac{dz'}{E(z')},
\end{equation}
where $H_0$ is the Hubble constant and 
\begin{equation}
E(z)=\sqrt{\Omega_{m}\, (1+z)^{3} + \Omega_{\Lambda}\, (1+z)^{3\, (1+w)}}
\end{equation}
is the normalized Hubble function for a flat $\rm \Lambda CDM$ universe. In this expression, $\Omega_{m}$ and $\Omega_{\Lambda}$ are the present-day matter and dark energy density in the unit of the critical density, respectively, and $w$ is the dark energy equation-of-state parameter. 

As the distance modulus $\mu$ is estimated after correcting for Galactic extinction, one expects an extinction overcorrection bias $\delta m(\phi) = - R_{\lambda}\, \delta E_{B-V}^{\rm EG}(\phi)$ (Equation~\ref{eq:ext_overcorrection}). This bias is redshift-dependent and can therefore impact cosmological parameter estimation. The estimated supernova dereddened distance modulus is, on average,
\begin{eqnarray}
    \langle \hat{\mu} \rangle_{\rm SN, dered}\, &=&\, \langle \mu \rangle_{\rm true} + \langle \delta m \rangle_{\rm SN}(z) \nonumber\\
    \, &=&\, 5\, \textrm{log} \left(\frac{\langle D_{\rm L}\rangle (z)}{10~\textrm{pc}}\right) - R_{\lambda}\, \langle \delta E_{B-V}^{\rm EG}\rangle_{\rm SN}(z)\,, 
\end{eqnarray}
where $\langle \delta E_{B-V}^{\rm EG}\rangle_{\rm SN}$, the excess reddening due to extragalactic imprints in the dust map, is to be evaluated at the zero lag ($\theta = 0$) toward the supernovae.

Here we investigate the fractional biases in $H_0$, $\Omega_m$, and $w$ under the level of extinction overcorrection effect measured in Section~\ref{sec:exgalresults}. For simplicity, we assume that the supernova hosts are galaxies similar to our SDSS reference objects, i.e., $\langle ... \rangle_{\rm SN} = \langle ... \rangle_{r}$. Based on the angular dependence of $ \langle \delta E_{B-V} \rangle_r$ we found that, as shown in Figure~\ref{fig:angular-correlation}, for maps with about 5 arcmin resolution like SFD and the $Planck$ maps, the zero lag $ \langle \delta E_{B-V} \rangle_r (\theta=0)$ is roughly 2 times the $10'$ averaged $\Delta E_{B-V}$ presented in Figure~\ref{fig:excess-ebv-redshift}. For the wavelength dependence $R_{\lambda}$, we adopt an extinction vector calibrated in \cite{2011ApJ...737..103S} assuming an $R_{V}$ of $3.1$ (this extinction vector includes a correction for a 13\% overestimation of the $E_{B-V}$ in SFD). Figure~\ref{fig:sn} shows the distance modulus offset $\delta \mu$ as a function of redshift after we correct for an extinction overcorrection for the SFD map in the rest-frame $B$ band (red data points). As is typically done in supernova cosmology studies, since there is an uncertainty in calibrating the distance ladder, we anchor the $\delta \mu$ to zero locally at our first redshift bin, $z\sim0.02$. Over a redshift range of order unity, extinction overcorrection thus changes the standard candles by 1.5 mmag. Figure~\ref{fig:sn} also overplots the distance modulus offset once we perturb each of the three cosmological parameters in each panel with the labeled fractional changes (black lines) from a fiducial flat universe cosmology of $[H_0,\ \Omega_m,\ w]=[70, 0.3, -1]$ (blue dash-dotted lines). We find that the bias given by the mean $\delta m$ in $H_0$ is small (less than $0.1\%$). For $\Omega_m$ and $w$ where a $0.5\%$ bias is expected, the effect in extinction overcorrection is going to be important when upcoming cosmology experiments are targeting $1\%$ level precision. Among these three cosmological parameters, accurate measurements of $w$ are of critical importance in the coming decade, as a significant departure of $w$ from unity would rule out the scenario of dark energy being a cosmological constant. We therefore suggest incorporating the correction of extinction-correction bias ($\sim0.5\%$) in upcoming Hubble diagram estimates.

\subsection{Bias Correction}\label{sec:correct_bias}

We now describe a procedure to correct for the biased dereddened magnitude estimations due to extragalactic imprints in Galactic dust maps. For a population of objects at a given redshift, the $10'$ average reddening excess $\Delta E_{B-V}$ presented in Figure~\ref{fig:excess-ebv-redshift} provides a starting point to quantify the amount of extinction overcorrection when a given dust map is used. Since the $\Delta E_{B-V}$ in Figure~\ref{fig:excess-ebv-redshift} is measured around specific types of reference objects (SDSS spectroscopic galaxies and quasars), for other types of objects, the mean excess reddening needs to be scaled with the clustering bias ratio. In Figure~\ref{fig:normalized_delta_ebv} we provide $\Delta E_{B-V}/b_r$, the excess-reddening estimations as a function of redshift for each map, similar to that present in Section~\ref{sec:exgalresults} but this time normalized by the bias of our reference sample (see Appendix B for the $b_r$ measurements).

With this information, we can determine the mean magnitude bias of an arbitrary galaxy population $g$ at redshift $z$ with a linear clustering bias $b_g$ following Equation~\ref{eq:pop_mag_shift}:
\begin{eqnarray}
\langle \delta m (z) \rangle_g \, &=&\, - R_{\lambda}\, \langle \delta E_{B-V} (z, \theta = 0) \rangle_g \nonumber\\
&=&\, -R_{\lambda}\, C\, \frac{\Delta E_{B-V}(z)}{b_r}\, b_g,
\end{eqnarray}
where we make clear that, since magnitude is a one-point statistic, the offset is related to the excess reddening at zero lag ($\theta = 0$). The constant $C$ is a beam correction to relate the $\Delta E_{B-V}$ measured within $10'$ to the zero-lag reddening excess. For maps like SFD or $Planck$ of about $5'$ resolution, $C\approx 2$, which can be visualized in Figure~\ref{fig:angular-correlation}; for lower-resolution maps with a spatial half width at half maximum compatible or larger than $10'$, $C\approx 1$ (thus, no beam correction is needed). For convenience, we provide a fitting function for the magnitude bias in the SFD map,
\begin{eqnarray}
\langle \delta m^{\textrm{SFD}}(z)\rangle_g\, &=&\, -R_{\lambda}\, \langle \delta E^{\textrm{SFD}}_{B-V}(z, \theta=0) \rangle_g \nonumber \\
&\simeq&\, -0.024\, R_{\lambda}\,b_g\,(z+0.16)^{-1.8}\;,
\end{eqnarray}
in mmag. To correct for this extinction overcorrection, one simply subtracts the $\langle \delta m \rangle$ (adds a positive magnitude) from the estimated dereddened magnitudes.

\section{Summary}\label{sec:summary}

We have analyzed 10 Galactic dust maps and investigated whether they are contaminated by extragalactic signals. Our tomographic analysis, based on the so-called clustering-redshift technique, has shown that 9/10 dust maps present imprints of extragalactic large-scale structure patterns, in some cases detected up to $z\sim4$. These extragalactic signals are found in all IR-based maps, from $12$ $\mu$m to the millimeter range, as well as ``stellar'' optical reddening maps. The amplitude of this extragalactic contamination is typically found to be at the millimagnitude level. Its redshift and angular scale variations depend on the chosen dust map. More specifically, we find the following.
\begin{enumerate}
    \item for all IR thermal dust maps, including the widely used \cite{1998ApJ...500..525S} map, Galactic reddening $E_{B-V}$ is systematically over-estimated around galaxies and quasars up to $z\sim2$ at a level ranging from subpercent to a few percent on scales of $10'$. This originates from CIB fluctuations due to the emission from dusty star-forming galaxies. The more recent \cite{2014A&A...571A..11P, 2016A&A...586A.132P} dust maps present a higher level of extragalactic contamination at higher redshift as they probe the Rayleigh-Jeans side of the dust blackbody emission spectrum, resulting in a negative $K$-correction. In addition, at $z>2$, this effect is further enhanced due to the peak of the cosmic star-formation history.

    \item For the stellar reddening maps using point-source optical photometry in Pan-STARRS1, we find an underestimation of Galactic reddening, especially around quasars at $1<z<1.5$ at the percent level. This reveals issues in  star--galaxy and/or star--quasar separations. 

    \item The $WISE$ 12$\mu$m map from \cite{2014ApJ...781....5M} is sensitive to PAH emission and has been used to create a Galactic dust map based on this tracer. Analyzing it, we detect the diffuse extragalactic PAH background up to $z\sim2$ and find that the Galactic PAH-to-dust ratio is similar to the cosmic mean.
    
    \item We have found the HI-based reddening map from \cite{2017ApJ...846...38L} to be free of extragalactic contamination, at least down to the $5\times10^{-4}$ level. Such a map provides an alternative to the more standard IR-based dust maps but relies on an assumed dust-to-gas ratio, whose spatial fluctuations can lead to an error of about 3 mmag in $E_{B-V}$.

\end{enumerate}
When these maps are used for correcting the photometry of extragalactic objects for Milky Way extinction, redshift- and scale-dependent biases are introduced. These artificial magnitude offsets then lead to biases in galaxy number counts and spatial auto- and cross-correlations at a level of about $10^{-3}$--$10^{-2}$ on scales of $10'$. These effects can then impact precision cosmology experiments. They can affect both object-based analyses and spatial statistics. These biases can be appreciable when estimating angular correlation functions with low amplitudes such as lensing-induced correlations or angular correlations for sources distributed over a broad redshift range. For precision cosmology with Type Ia supernovae, we expect a $0.5\%$ impact on the determinations of $\Omega_m$ and $w$, which will be significant for upcoming surveys like LSST and WFIRST targeting the one percent precision range. For such experiments, we recommend testing the robustness of the final results against the different dust maps used. Finally, we provide a procedure to correct for or decrease the level of biased magnitude corrections in maps with extragalactic imprints.

\begin{acknowledgements}
Y.C. acknowledges support from NSF grant AST1313302 and NASA grant NNX16AF64G. We thank Aaron Meisner, Brandon Hensley, Bruce Draine, Daniel Lenz, David Schlegel, Douglas Finkbeiner, Edward Schlafly, Francois Boulanger, Joshua Peek, Marc-Antoine Miville-Deschenes, Mathieu Remazeilles, and Olivier Dor{\'e} for helpful discussions.

\end{acknowledgements}

\bibliographystyle{apj}

\appendix

\section{A.Error Maps, Optimal Weighting, and Reflattening}

Here we describe the technical details of the excess-reddening estimator $\langle \delta E_{B-V}(\theta, z) \rangle_r$ defined in Equation~\ref{eq:cross-correlation} in Section~\ref{sec:formalism}, where the reddening fluctuation field $\delta E_{B-V}(\phi)$ is defined in Equation~\ref{eq:delta_ebv} with a zero point taken from a one degree running mean. If the signal-to-noise ratio is roughly constant on the sky, a simple mean for the $\langle ... \rangle$ operation is already optimal. In our case, however, the signal-to-noise ratio has a strong spatial gradient: while the signal (extragalactic imprint) stays roughly constant on large scales, the Galactic foreground varies strongly across the sky even at high latitudes. We therefore adopt an inverse variance-weighted mean for the $\langle ... \rangle$ operation, and the excess-reddening estimator becomes 
\begin{equation}
    \langle \delta E_{B-V}(\theta, z) \rangle^{\rm wgt}_r \equiv  \frac{\langle W (\theta) \delta E_{B-V}(\theta, z) \rangle_r}{\langle  W(\theta) \rangle_r},
\end{equation}
where $W(\phi)$ is the weight field derived from taking the squared inverse of an error map of the $\delta E_{B-V}(\phi)$ field. We take an empirical approach to estimate such an error map of one degree resolution. Within any $1^{\circ}$ (or larger) patch on the sky, we can evaluate the distribution function of the $\delta E_{B-V}(\phi)$ values: the scatter in this distribution due to extragalactic signal is expected to be roughly invariant in different patches on the sky, but the observed scatter can vary strongly due to variations in the Milky Way foreground or photon noise due to nonuniform scanning patterns. We thus calculate a $1^{\circ}$ running $68$th percentile scatter field of the $\delta E_{B-V}(\phi)$ and take it as our error map and its squared inverse as the weight map $W(\phi)$. We found that in our case, this weighting scheme improves the signal-to-noise ratio of our $\langle \delta E_{B-V}(\theta, z) \rangle_r$ estimator by a factor of about two, which is essential for some of the weak features presented in Section~\ref{sec:exgalresults}. To ensure a flat zero point on large scale using this weighting scheme, we also ``reflatten'' the $\delta E_{B-V}(\phi)$ field in Equation~4 by subtracting its $1^{\circ}$ running weighted mean. This is a small correction but guarantees removal of spurious cross-correlation results due to a drifting zero point.

\section{B. Reference Sample Clustering and Its Modulation to Reference-reddening Correlations}

The clustering properties of the reference sample can modulate the amplitudes of the reference--dust map correlation presented in Section~\ref{sec:exgalresults}. To investigate this effect, we estimate the biases of the SDSS MAIN, LOWZ, and CMASS samples by measuring their autocorrelation functions and comparing with that expected for dark matter in our assumed cosmology using the CLASS code \citep{2011arXiv1104.2932L}. For SDSS quasars, we use an analytic function from \cite{2006MNRAS.371.1824P} fitted to the measurements in \cite{2006MNRAS.371.1824P}, \cite{2007AJ....133.2222S}, and \cite{2015MNRAS.453.2779E}.

\begin{figure}[t]
    \begin{center}
         \includegraphics[width=0.45\textwidth]{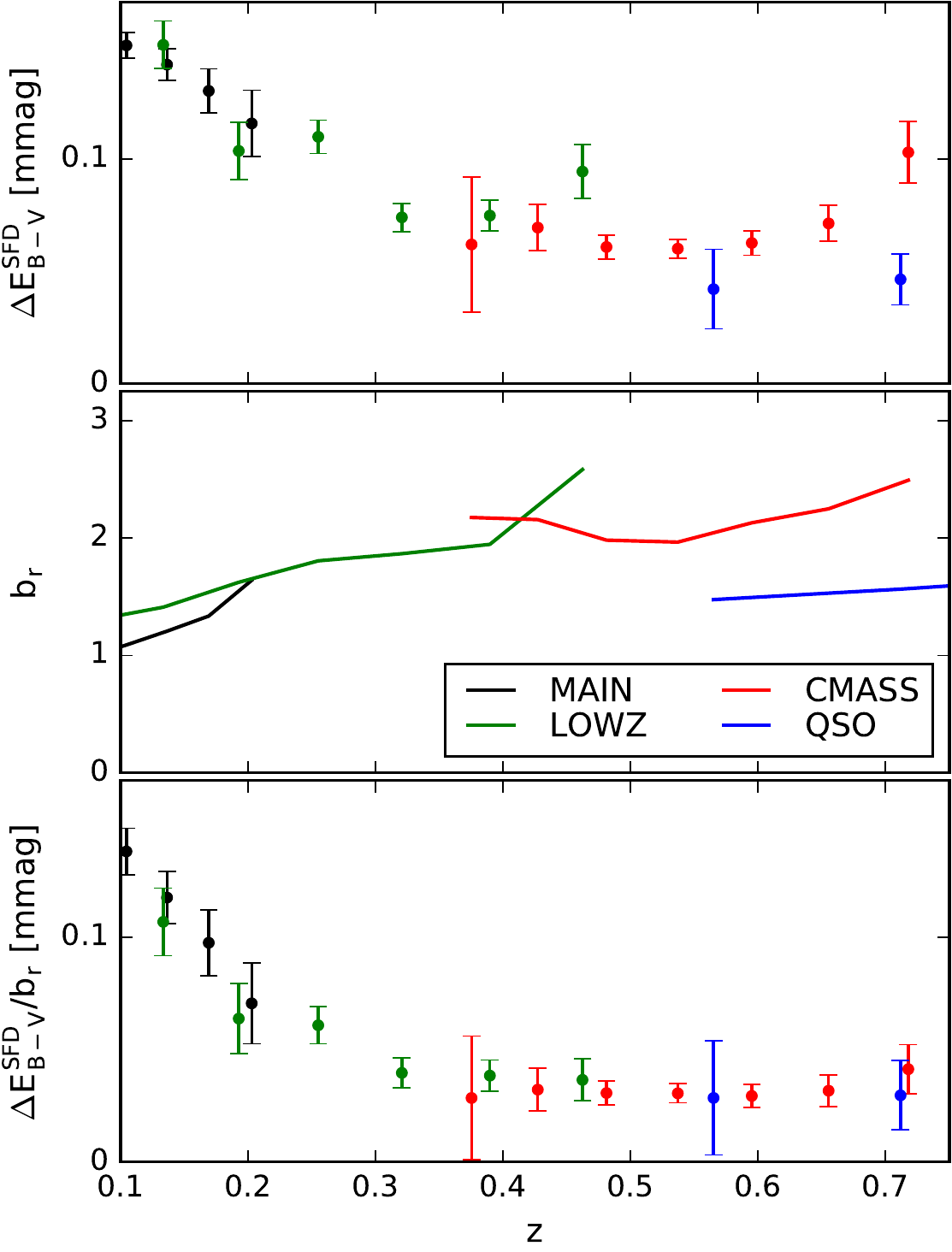}
    \end{center}
    \caption{Top panel: excess $E_{B-V}^{\rm SFD}$ around each reference sample. Some differences are found at redshift intervals where multiple reference samples are available. Middle panel: linear galaxy bias for each reference sample obtained via angular autocorrelation. Bottom panel: bias-corrected excess $E_{B-V}^{\rm SFD}$, in which all the measurements are consistent with being drawn from the same distribution. This suggests that the extragalactic imprint can be considered as yet another tracer of matter density field, insensitive to which reference sample is used to extract it.}
    \label{fig:bias-correction}
\end{figure}

In Figure~\ref{fig:bias-correction} we demonstrate that our extragalactic imprint estimation $\Delta E_{B-V}$ (Equation~\ref{eq:excess-reddening-redshift}) indeed scales linearly with the clustering bias of the reference sample. The top, middle, and bottom panels of Figure~\ref{fig:bias-correction} show the excess reddening in SFD around each reference sample, the linear bias of each reference sample, and the bias-corrected excess reddening, respectively. Before the bias correction, some differences in $\Delta E_{B-V}$ are seen over the redshift intervals at which we have multiple reference samples, while the linear bias normalization brings all the measurements to a unique redshift dependence. At $z\sim 0.6$--$0.7$, where we have both CMASS LRGs and quasars, if the excess reddening was dominated by emission of the reference objects themselves, one would expect a lower signal around LRGs, since they have little dust, ongoing star-formation, and thus far-IR fluxes. Instead, we find the opposite: the $\Delta E_{B-V}$ measured around LRGs is higher than that around quasars with a factor consistent with the bias ratio. The extragalactic imprints we detect thus should thus be understood as the far-IR emission from galaxies in the large-scale structure tracing the underlying matter density field (as opposed to the star-formation of the reference objects themselves). The reference bias-corrected $\Delta E_{B-V}$ over the entire redshift range for all the dust maps considered is provided in Figure~\ref{fig:normalized_delta_ebv} using the combined $b_r$ (number weighted over all four reference samples) measured here.

\section{C. Effect of Dust Correction in the Reference Sample}
The density field of the dereddened SDSS reference sample could already be modulated by the extinction overcorrection bias (in SFD, the map used by the SDSS collaboration) according to Equation~\ref{eq:apparent_delta_g}. Here we show that this does not significantly affect our cross-correlation measurements of dust map extragalactic imprints at the amplitudes currently measured in Section~\ref{sec:exgalresults}. Keeping the extinction-correction bias in the density contrast of reference objects, our reference-reddening excess correlation estimator (Equation~\ref{eq:cross-correlation}) becomes
\begin{eqnarray}
    \langle \hat{\delta}_{r,{\rm dered}} \cdot \delta E_{B-V} \rangle \, &=&\, \langle \delta E_{B-V}^{\rm EG} \rangle_r + \alpha\, \langle \delta \mu^{\rm SFD} \cdot \delta E_{B-V} \rangle \nonumber \\
    &=&\, \langle \delta E_{B-V}^{\rm EG} \rangle_r + 0.92\, \alpha\,  R_{\lambda} \, \langle \delta E_{B-V}^{\rm EG, SFD} \cdot \delta E_{B-V}^{\rm EG} \rangle\, , \nonumber \\
    \label{eq:excess-reddening-revisit}
\end{eqnarray}
where the extra term is a cross-correlation between the extragalactic imprints in two dust maps, one used in dereddening the reference sample (assuming SFD) and one currently being tested for extracting potential extragalactic imprints. This term, although being still purely extragalactic, would complicate the interpretation of $\langle \hat{\delta}_{r,{\rm dered}} \cdot \delta E_{B-V} \rangle$ being a simple galaxy--reddening correlation. We can set an upper limit of $\langle \delta E_{B-V}^{\rm EG, SFD} \cdot \delta E_{B-V}^{\rm EG} \rangle$ by measuring $\langle \delta E_{B-V}^{\rm SFD} \cdot \delta E_{B-V} \rangle$, i.e., the total power in the reddening--reddening correlation, which we find it to be of order $10^{-5}$ at zero lag ($\theta = 0$). This is much lower than the first term $\langle \delta E_{B-V}^{\rm EG} \rangle_r$ of order $\sim 10^{-3}$ and most of the error bars presented in Section~\ref{sec:exgalresults}. In other words, the true 3D clustering of the reference galaxies (order of above unity at relevant scales) is much stronger than the extinction overcorrection bias; thus, the dust map bias autocorrelation becomes negligible in this galaxy-reddening correlation estimator.

\end{document}